\documentclass[aps,pre,twocolumn,superscriptaddress,showpacs]{revtex4-1}% Physical Review 
\usepackage[pdftex]{graphicx}% Include figure files
\pdfoutput=1

\usepackage{graphicx}% Include figure files
\usepackage{dcolumn}% Align table columns on decimal point
\usepackage{bm}% bold math
\usepackage{amssymb}
\makeatletter
\def\captionof#1#2{{\def\@captype{#1}#2}}
\makeatother

\usepackage[english]{babel}
\usepackage{times}
\usepackage{graphicx}
\usepackage{amsfonts}
\usepackage{psfrag}
\usepackage{verbatim}
\usepackage{color}
\begin{document}

%opening
\title{Dynamics of a homogeneous active dumbbell system}

\author{Antonio Suma} 
\affiliation{SISSA - Scuola Internazionale Superiore di Studi Avanzati,\\
Via Bonomea 265, 34136 Trieste 
Italy}
\email{antonio.suma@sissa.it}

\author{Giuseppe Gonnella} 
\affiliation{Dipartimento di Fisica, Universit\`a di Bari {\rm and}  \\
INFN, Sezione di Bari, via Amendola 173, Bari, I-70126,
Italy}
\email{gonnella@ba.infn.it}

\author{Gianluca Laghezza} 
\affiliation{Dipartimento di Fisica, Universit\`a di Bari {\rm and}  \\
INFN, Sezione di Bari, via Amendola 173, Bari, I-70126,
Italy}
\affiliation{Current address: Department of Physics, Theoretical Physics, University of Oxford,\\
1 Keble Road, Oxford OX1 3NP, United Kingdom
}
\email{gianluca.laghezza@physics.ox.ac.uk}

\author{Antonio Lamura}
\affiliation{Istituto Applicazioni Calcolo, CNR, via Amendola 122/D,\\
Bari, I-70126, Italy} 
\email{a.lamura@ba.iac.cnr.it}
   
\author{Alessandro Mossa} 
\affiliation{Dipartimento di Fisica, Universit\`a di Bari {\rm and}  \\
INFN, Sezione di Bari, via Amendola 173, Bari, I-70126,
Italy}
\email{Alessandro.Mossa@ba.infn.it}

\author{Leticia F. Cugliandolo} 
\affiliation{Sorbonne Universit\'es, Universit\'e Pierre et Marie Curie
  - Paris 6, Laboratoire de Physique Th\'eorique et Hautes Energies,
 \\ 4, Place Jussieu, Tour 13, 5\`eme \'etage, 75252 Paris Cedex 05,  France}
\email{leticia@lpthe.jussieu.fr}

\begin{abstract}
We analyse the dynamics of a two dimensional system of interacting active dumbbells. We characterise the mean-square displacement, linear 
response function and deviation from the equilibrium fluctuation-dissipation theorem as a function of activity strength,
packing fraction and temperature for parameters such that the system is in its homogeneous phase. While the diffusion constant 
in the last diffusive regime naturally increases with activity and decreases with packing fraction, we exhibit
an intriguing non-monotonic dependence on the activity  of the ratio between the finite density and the single particle 
diffusion constants. At fixed packing fraction, the time-integrated linear response function depends non-monotonically 
on activity strength.
The effective temperature extracted from the ratio between the integrated linear response and 
the mean-square displacement in the last diffusive regime is always higher than the ambient temperature, increases with 
increasing activity and, for small active force it  monotonically increases with density while for sufficiently high activity 
it first increases to next decrease with the packing fraction. We ascribe this peculiar effect to the existence of finite-size 
clusters for sufficiently high activity and density at the fixed (low) temperatures at which we worked. The crossover occurs 
at lower activity or density the lower the external temperature. The finite density effective temperature is higher (lower) than
the single dumbbell one below (above) a cross-over value of the P\'eclet number. 
\end{abstract}

%Use showkeys class option if keyword

\maketitle

\setlength{\textfloatsep}{10pt} 
\setlength{\intextsep}{10pt}

%%%%%%%%%%%%%%%%%%%%%%%%%%%%%%%%%%%%%%%%%%%%%
%\section{Introduction}
%%%%%%%%%%%%%%%%%%%%%%%%%%%%%%%%%%%%%%%%%%%%%

\section{Introduction} 

Active matter is constituted by self-propelled units  that extract
energy from internal sources or its surroundings, and are also in contact with an environment
that allows for dissipation and provides thermal fluctuations. 
The locally gained energy is partially converted into work and 
partially dissipated into the bath. The units can interact 
via potential forces or through disturbances in the medium. This new class of (soft) matter is the focus of 
intense experimental,
theoretical and numerical studies for practical as well as fundamental reasons. Several review articles summarise the current understanding of 
active systems~\cite{Toner05,Fletcher09,Menon10,Ramaswamy10,Cates12,Vicsek12,Marchetti13,Safran13,Marenduzzo14}.

Due to the consumed energy, detailed balance is broken in active matter, and these systems are 
inherently out of equilibrium. Natural examples  
are bird flocks, schools of fish, and bacterial colonies. Artificial self-propelled particles have also been realised in the laboratory
by using, for instance, granular materials~\cite{Narayan07,Deseigne10} or colloidal particles with specific surface treatments~\cite{Paxton04,Hong06,Palacci10} 
and are especially suited for  experimental tests. Different models  have been proposed to mimic these systems.
For instance, run-and-tumble motion is used to model {\it Escherichia coli} bacteria~\cite{Cates12} while active Brownian particles are 
used to model Janus colloidal particles~\cite{Romanczuk12}. 

Self-propulsion is responsible for  many interesting, and also sometimes
surprising, collective phenomena. 
Some of these are: the existence of orientationally ordered states in two spatial dimensions~\cite{Czirok97,vicsek},
spatial phase separation into an aggregated phase and gas-like regions for sufficiently large 
packing fractions in the absence of attractive interactions~\cite{Tailleur08,Fily12,Fily14,Redner13,Buttinoni13b,Stenhammer13,Suma13,Suma14,Levis-Berthier,Stenhammer14}, 
giant density fluctuations~\cite{Ramaswamy03,Ginelli10,Fily12,Buttinoni13b} and 
accumulation at boundaries~\cite{Elgeti09,Elgeti13,Bricard13}, spontaneous collective motion~\cite{Wang11b}, glassy 
features~\cite{Shen04,Angelini11,Henkes11,Berthier-Kurchan},
unexpected rheological properties~\cite{Chate06} and non-trivial behavior under shear~\cite{Cates08,Saracco09,Bees10,Bearon12,Saracco12}.

Active systems, being essentially out of equilibrium, also pose many  fundamental physics questions such as 
whether thermodynamic concepts could apply to them in their original setting or with simple modifications.

The {\it effective temperature} notion was proposed to describe some macroscopic aspects of slowly relaxing macroscopic 
physical systems, such as glassy systems and  gently sheared super-cooled liquids~\cite{cugl-kur-pel,cugl:review}.
This (potentially) thermodynamic  intensive parameter is defined as the parameter replacing ambient temperature in the 
(multi) time dependent fluctuation-dissipation relations between induced and spontaneous out of equilibrium fluctuations
of the system. 
In systems with multiple time-scales special care has to be taken in the choice of the time-regime 
in which a thermodynamic-like parameter could be extracted. More precisely, experience shows that one may identify it 
in the time-regime of (large) structural relaxation, while at  short-time scales  the microscopic dynamics imposes the 
system's fluctuations (be them quantum, active or thermal). To retain a thermodynamic sense, the effective temperature 
should also be measurable with suitable choices of thermometers such as well-chosen tracer particles and
it should be the same for all observables evolving in the same time regime.

The effective temperature idea has been explored, to a certain extent, in the context of active matter.
The effective temperature of a bacterial bath was estimated from the Stokes-Einstein relation of a 
tracer particle in~\cite{wu}. The effective temperature notion was used to characterise crystallization effects known to occur under 
large active forces~\cite{Shen04,Shen05,Wang11} and the emergence of collective motion~\cite{Wang11b}. The deviations from the 
equilibrium fluctuation-dissipation theorem in equilibrium were used to reveal  the active process in 
hair bundles~\cite{Martin01} and model cells~\cite{Mizuno07}.
In biological systems the nature of the microscopic active elements is difficult to study directly.
The fluctuation-dissipation relations could be useful to characterise 
the active forces in active matter in 
general, and in living cells in particular. With this idea in mind, Ben-Isaac {\it et al.} analysed the perturbed and 
spontaneous dynamics
of blood cells in the lab, and compared the outcomes to the ones of a single particle Langevin model with a special choice of the statistics of the active 
forces analytically~\cite{Ben-Isaac-etal}.  Similar ideas were used in~\cite{Bohec13,Fodor14} 
to characterise motor activity in living cells and actin-myosin networks.

We stressed the fact that the effective temperature is not a static parameter in the sense that it cannot be read from a system's snapshot. 
It should be determined from dynamic measurements in 
which the separation of time-scales has to be very carefully taken into account to obtain sensible results~\cite{cugl-kur-pel,cugl:review}.
Having said this, the effective temperature has been found to play a role similar to ambient temperature in the celebrated  experiment
of Perrin
now performed with active particles. Indeed, 
the sedimentation of an active colloidal suspension of Janus particles under a gravity field 
exhibits the same exponential density distribution as a standard
dilute thermal colloidal system. The only difference is that the parameter
that replaces the thermal system's temperature in the active case is equal to
 the effective temperature  inferred from an
independent measurement of the long-time diffusive motion of an active colloidal particle~\cite{Palacci10}. 
This problem was studied analytically with a run-and-tumble model~\cite{Tailleur09}  and a Langevin process  for a 
tagged active point-like particle with a suitable choice of activity~\cite{Szamel14}.

Deviations from the  equilibrium fluctuation-dissipation 
theorem, as well as other ways of measuring the effective temperature by using tracers,
were analysed numerically by Loi {\it et al.} in (relatively loose) systems of active 
point-like particles~\cite{cugl-mossa1,cugl-mossa3} and long molecules~\cite{cugl-mossa2,cugl-mossa3}.
All these measurements yielded consistent results.
In this paper, we will follow this kind of analysis in a model of active matter that we now discuss.

In their simplest realisation, active units are taken to be point-like. However, 
the importance of the shape and polarity of the self-propelled particles for their collective behaviour has been 
stressed in the literature~\cite{Peruani06,yang2010swarm, Peruani12prl, Baskaran12, Wensink12}.
Active units, whether synthetic or natural, are typically rodlike or elongated. This is the case of most bacteria, 
chemically propelled 
nanorods, and actin filaments walking on molecular motor carpets. The length-scale of these units is of the order
of several micrometers.

A simple way to model a {\it shortly} elongated 
swimmer is to use a dumbbell, consisting of two colloids linked by a Hookean spring (see~\cite{Wensink} and references therein). 
Such passive dumbbell models have been used to mimic the viscolelastic  behavior of 
linear flexible polymers suspended in  Newtonian fluids~\cite{Espanol09, Winkler13}.
Hydrodynamic interactions between dumbbell swimmers have been considered in~\cite{Yeomans08,Yeomans10,Gyrya}. 
In this study we add activity to Brownian dumbbells in the form of a constant
propulsive force acting on the direction connecting the two beads. We include potential 
interactions between the dumbbells but we do not impose any alignment rule.
This model was used to describe bacterial systems~\cite{valeriani2011colloids}. 
Compared with self-propelled spherical particle models, it phase separates at smaller  densities~\cite{Suma13,Suma14}.
Clustering and phase separation are here due
to the out of equilibrium drive exerted by the persistent local
energy input that breaks detailed balance. Moreover, together with  spontaneous  aggregation, 
dumbbells
break chiral invariance, and rotate
displaying nematic order with spiral patterns.

We focus on the dynamic behaviour 
in the  homogeneous phase of the two dimensional system. 
We study its dynamic properties at various fixed temperatures and dumbbell parameters as will be introduced below, 
but  for a broad range of values of the surface density and strength of the active force. We analyse 
the behaviour of the single passive and active molecule analytically and we later use molecular dynamics 
simulations to study the 
many-body system. More precisely, we compute the mean-square displacement, linear response function, their relation and 
its implications on the effective temperature ideas~\cite{cugl:review} that we discuss in detail in the main text.

The body of the paper is organised as follows. In Sec.~\ref{subsec:model-dumbbells} we introduce the model.
In Section~\ref{sec:algorithm} we give some details on the numerical algorithm that we use to 
study the problem numerically. Section~\ref{sec:passive} is devoted to the study of the  passive dumbbell system, both in its 
single molecule limit and many-body case. In Sec.~\ref{sec:single-active} we present the analysis of a single active dumbbell molecule and
in Sec.~\ref{sec:active-many-body} the one of a system of interacting and active dumbbell molecules.
Finally, in Sec.~\ref{sec:conclusions} we summarise our results and present our conclusions.

%\section{Dumbbell system}

%In this Section we introduce the dumbbell model and we
%briefly discuss its dynamics  in the absence of tracers. The Section is organized as follows.
%We first give some details on the algorithm and parameters
%used in the simulations. We then review the behaviour of the passive dumbbell 
%system, paying special attention to the dependence of
%the dynamic regimes on the surface fraction $\phi$.
%We turn next to the active case, where we discuss the dynamics of a single 
%molecule and the ones of an interacting system. This section sets the arena to the use of kinetic and 
%potential tracers to investigate the dynamics of the active dumbbell system, the problem 
%to be attacked in Sec.~\ref{sec:tracers-in-dumbbell-system}.

\section{The model}
\label{subsec:model-dumbbells}

A dumbbell is a diatomic molecule made of two spherical colloids with diameter $\sigma_{\rm d}$ 
connected by a spring of elastic constant $k$ 
that one can mimic, in its simplest form, with 
Hooke's law
\begin{equation}
V_{\rm H}( r  ) = \frac{1}{2} k r^2
\; ,
\label{eq:hook}
\end{equation}
with $r$ the distance between their centers of mass.
 An additional repulsive force is added, derived from just the  repulsive part
of a Lennard-Jones potential, that ensures that the two colloids cannot overlap. This 
potential is called Weeks-Chandler-Anderson (WCA) and it is given by~\cite{Weeks} 
\begin{eqnarray}
\label{eq:WCA-potential}
V_{\rm wca}( r ) 
&=&
\left\{
\begin{array}{ll}
V_{\rm LJ}( r ) - V_{\rm LJ}(r_c) & \qquad r<r_c
\nonumber\\
0 & \qquad r > r_c
\end{array}
\right.
\end{eqnarray}
with 
\begin{equation}
V_{\rm LJ}(  r  ) = 4\epsilon \left[ \left( \frac{\sigma_{\rm d}}{r} \right)^{12} - \left( \frac{\sigma_{\rm d}}{r} \right)^{6}\right]
\; .
\end{equation}
$\epsilon$ is an energy scale and $\sigma_{\rm d}$ is, once again, the diameter of the spheres 
in the dumbbell. $r_c$ is the minimum of the Lennard-Jones potential, $r_c=2^{1/6} \sigma_{\rm d}$.
We neglect hydrodynamic interactions.

The equation of motion for a single dumbbell immersed in a liquid is the Langevin equation 
\begin{eqnarray}
m_{\rm d} \ddot {\mathbf r}_i &=& -\gamma \dot {\mathbf r}_i - k ({\mathbf r}_i- {\mathbf r}_j)
\nonumber\\
&& 
- \frac{\partial V_{\rm wca}(r_{ij})}{\partial r_{ij}} \frac{({\mathbf r}_i - {\mathbf r}_j)}{r_{ij}} +{\boldsymbol \eta}_i
\label{eq:Langevin}
\end{eqnarray}
with $i,j=1,2$ labelling the two spheres in the molecule,
${\mathbf r}_i$ the position of the $i$-th monomer with respect to the 
origin of a Cartesian system of coordinates fixed to the laboratory, 
${\mathbf r}_{ij} = {\mathbf r}_i - {\mathbf r}_j$ and $r_{ij} = |{\mathbf r}_{ij}|$.
$m_{\rm d}$ is the mass of each sphere in the dumbbell and $\gamma$ is the friction 
coefficient. 
The Gaussian noise has zero mean and it is delta-correlated
\begin{eqnarray}
\langle \eta_{ia}(t) \rangle &=& 0 \; , 
\label{eq:noise-ave}\\
\langle \eta_{ia}(t) \eta_{jb}(t') \rangle &=& 2 \gamma k_BT \delta_{ij} \delta_{ab} \delta(t-t')
\; ,
\label{eq:noise-corr}
\end{eqnarray}
with $k_B$ the Boltzmann constant and $T$ the temperature of an equilibrium environment in which the 
dumbbells move. $a, b$ label the coordinates in $d$ dimensional space. Note that an effective 
rotational motion is generated by the random torque due to the white noise acting independently on the two beads.

We add now the active force to Eq.~(\ref{eq:Langevin}). It acts
in the direction of the spring linking the two 
colloids, i.e. in the direction $\hat {\mathbf n}$
of the straight line passing by the two centers of mass,
and it is constant in modulus. It reads 
\begin{equation}
{\mathbf F}_{\rm act} = F_{\rm act} \ \hat {\mathbf n}
\; . 
\end{equation}

Having established the single molecule stochastic model, we 
extend it to consider a system of $N$ such bi-atomic molecules in interaction
immersed in a bidimensional space with surface $S$. The molecule number density is  
$n_{\rm d}=N/S$, 
%the mass density is $\rho=2m_{\rm d}N/S$ 
and the surface fraction is 
\begin{equation}
\phi=N \ \frac{S_{\rm d}}{S} 
\;  
\end{equation}
with $S_{\rm d}$ the area occupied by an individual dumbbell. The spring is supposed to be massless and void of surface. Therefore, in  
$d=2$ we have $S_{\rm d} = \pi \sigma_{\rm d}^2/2$.  
 
In order to model the many-body system we introduce an inter-molecular potential and we 
slightly modify the interaction between the spheres in the same molecule.
The interaction between dumbbells is purely repulsive and avoids the 
superposition of different molecules, i.e. ensures the 
excluded volume condition  in the WCA way. We use then a Lennard-Jones potential 
still truncated to have only the repulsive part as in Eq.~(\ref{eq:WCA-potential}). The elastic 
potential between spheres belonging to the same dumbbell is modified to be 
of the finite extensible non-linear elastic (FENE) kind  to avoid the unlimited 
separation of the colloids belonging to the same molecule:
\begin{equation}
{\mathbf F}_{\rm fene} = \frac{k {\mathbf r}}{1- (r^2/r_0^2)} 
\; . 
\end{equation}
The denominator ensures that the spheres cannot go beyond the distance $r_0$.

\begin{figure}[h]
\includegraphics[scale=0.3]{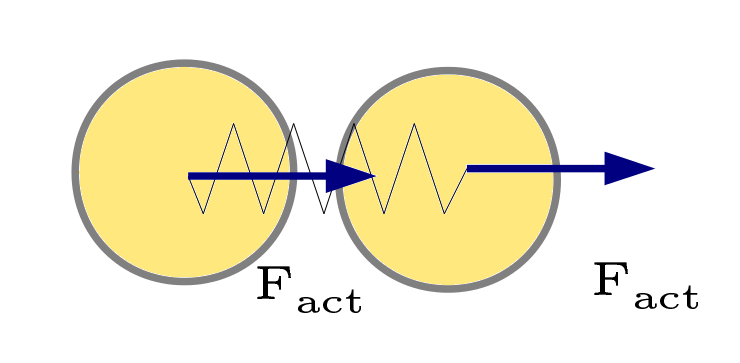}
\caption{(Color online.) A sketch of an active dumbbell molecule.}
\label{fig:dumbbell-act}
\end{figure}

The dynamic equations for one dumbbell in the system are
\begin{eqnarray}
 m_d\ddot{{\mathbf r}}_{i}(t) &=& -\gamma \dot{{\mathbf r}}_{i}(t)-
 {\mathbf F}_{\rm fene}({\mathbf r}_{i, i+1})+
 {\boldsymbol \eta}_{i} 
 \nonumber\\
 &&
 -\sum_{j=0,j\neq i}^{2N} 
 \frac{\partial V_{\rm wca}^{ij}}{\partial  r_{ij}}
 \frac{{\mathbf r}_{ij}}{ r_{ij}}+ {{\mathbf F}_{\rm act}}_i
\; , \\
 m_d\ddot{{\mathbf r}}_{i+1}(t) &=& 
 -\gamma \dot{{\mathbf r}}_{i+1}(t)+
 {\mathbf F}_{\rm fene}({\mathbf r}_{i,i+1})+
 {\boldsymbol \eta}_{i+1}
 \nonumber\\
 &&
 -\sum_{j=0,j\neq i+1}^{2N} 
 \frac{\partial V_{\rm wca}^{i+1,j}}{\partial r_{i+1,j}}
 \frac{{\mathbf r}_{i+1,j}}{r_{i+1,j}}+{{\mathbf F}_{\rm act}}_i \; ,  
 \label{eqdumbattcoll}
\end{eqnarray}
with $i=1,3,...2N-1$
 and $V_{\rm wca}^{ij} \equiv
V_{\rm wca}(r_{ij})$ with $V_{\rm wca}$ defined in Eq.~(\ref{eq:WCA-potential}).
The statistics of the noise ${\boldsymbol \eta}$ is Gaussian with average and correlation 
given by Eqs.~(\ref{eq:noise-ave}) and (\ref{eq:noise-corr}), respectively. As the 
active force's direction lies along the molecular axis it depends on the diatomic molecule
but is the same for the two atoms. This is the reason why we label it $i$ in the 
two equations above.  Note that once the active force is attached to a molecule 
a sense of back and forth atoms is attributed to them (see Fig.~\ref{fig:dumbbell-act}). The active forces are  applied to all molecules
in the sample during all their dynamic evolution. 
${\mathbf F}_{\rm act}$ is a time-dependent vector since, although its modulus is 
constant, it does change direction 
together with the molecule's rotation. For each dumbbell ${\mathbf F}_{\rm act}$ is  directed 
from the colloid $i$ (tail) to the  colloid $i+1$ (head). Note the difference between this 
kind of activity and the random one used in~\cite{cugl-mossa2,cugl-mossa3} for the numerical study of 
active polymers.

The P\'eclet number, ${\rm Pe}$, is a dimensionless ratio between the  
advective transport rate and the diffusive transport rate.
For particle flow one defines it as 
%\begin{equation}
$
%{\rm Pe} = \frac{Lv}{D} 
{\rm Pe} = Lv/{D} 
$
%\;
 , 
%\end{equation}
with $L$ a typical length, $v$ a typical velocity, and $D$ a typical diffusion constant
in the problem. We choose $L \to \sigma_{\rm d}$, $v\to F_{\rm act}/\gamma$ and 
$D\to D^{\rm pd}_{\rm cm}= k_BT/(2\gamma)$ of the passive dumbbell [see Eq.~(\ref{eq:pddiffusion})]; then,
\begin{equation}
{\rm Pe} = \frac{2\sigma_{\rm d} F_{\rm act}}{k_BT}
\; . 
\label{eq:Peclet}
\end{equation}
Another important parameter is the active Reynolds number 
\begin{equation}
{\rm Re}_{\rm act} 
= \frac{m F_{\rm act} }{\sigma_{\rm d} \gamma^2} 
\; , 
\label{eq:Reynolds}
\end{equation}
defined in analogy with the usual hydrodynamic Reynolds number ${\rm Re}=  L v /\nu$,
where $\nu $ is the kinematic viscosity of a given fluid, representing the ratio between inertial
and viscous forces. Here we set
 $L \to \sigma_{\rm d}$, $v\to F_{\rm act}/\gamma$ and 
$\nu \to \gamma \sigma_{\rm d}^2 /m_{\rm d}$.

%In our simulations we always choose values of Fa such
%that Rea << 1 in analogy with the Stokes regime.

\section{The molecular dynamics algorithm}
\label{sec:algorithm}

We solved the stochastic Langevin equation with an algorithm due to 
Vanden-Eijinden and Ciccotti~\cite{Cicotti} 
that is exact to order $(\Delta t)^2$ with $\Delta t$ the time-step. 
We used a square bidimensional box 
%of size $l$ 
with periodic boundary conditions.

The units of mass, length and energy are $m_{\rm d}$, $\sigma_{\rm d}$, and $\epsilon$, respectively.
The adimensional elastic constant is defined as $k^*=k\sigma_{\rm d}^2/\epsilon$
and we take a large value $k^*=30$ to avoid the excessive extension of the 
dumbbell, and also prevent vibrations.
 The length $r_0$ in the FENE potential is rendered adimensional 
as $r_0^* = r_0/\sigma_{\rm d}$ and we used $r_0^*=1.5$.  The thermal energy  
is normalized by the energy scale in the Lennard-Jones potential, $k_BT^*=k_BT/\epsilon$.
The friction constant has dimensions of 
mass/time and we can make it adimensional as 
$\gamma_{\rm d}^*=\gamma/\sqrt{\epsilon m_{\rm d} /\sigma_{\rm d}^2}$. 
Physical realizations are usually in the over-damped regime which is ensured 
 by choosing a  value of $\gamma_{\rm d}^*$ such that  ${\rm Re}_{\rm act} \ll 1$.
%$\gamma > 2\sqrt{m_{\rm d}k}$ that, in adimensional parameters reads
%$\gamma_{\rm d}^* > 2 \sqrt{k^*}$
We preferred to consider a model with inertial terms in order to have access to the  
 crossover between  ballistic and diffusive regimes.
Concretely, we used $\gamma_{\rm d}^*=10$ for which the molecular  oscillations are  strongly inhibited.
Finally, the adimensional temperatures $k_BT^*$ are in 
the range $0.001 - 1$. Once all parameters are expressed in 
terms of reduced units we can effectively set $m_{\rm d}=\sigma_{\rm d}=k_B= \epsilon=1$.

The optimal choice of time-step is delicate. We first identified some relevant time-scales
in the problem and we later chose $\Delta t$. The time-scale for the oscillations 
of the free dumbbell is  $\tau_{\rm osc} = 2\pi \sqrt{m_{\rm d}/k} = 
2\pi/\sqrt{k^*} \sqrt{m_{\rm d}\sigma_{\rm d}^2/\epsilon}$ (and equals 1.14 for our choice of parameters). The inertial time scale is 
$t_I = m_{\rm d}/\gamma = 1/\gamma_d^* \ \sqrt{m_{\rm d}\sigma_{\rm d}^2/\epsilon}$ (and equals 0.1 for our choice of parameters).
We will show below that there is another time-scale, typically longer, associated to the angular diffusion, 
$t_a=\gamma R_\infty^2/(2k_BT)\approx \gamma_{\rm d}^*/(2k_BT^*)
\sqrt{m_{\rm d} \sigma_{\rm d}^2/\epsilon}$ where we used $R_\infty\approx \sigma_{\rm d}$ the typical length 
of the dumbbell molecule. We chose to work with 
$\Delta t = t_I/100$ to see details of the ballistic regime and $\Delta t = \tau_{\rm osc}/100$ to enter the 
later diffusive and active regimes. Times are measured in units of $\sqrt{(\sigma_{\rm d}^2 m_{\rm d})/\epsilon}$.

\section{Passive system}
\label{sec:passive}

In this Section we review the behaviour of the passive dumbbell 
single molecule and interacting system, paying special attention to the dependence of
the dynamic regimes on the surface fraction $\phi$.

\subsection{Passive single dumbbell}

One can simply show that the equation
of motion for the position of the center-of-mass, ${\mathbf r}_{\rm cm}= ( {\mathbf r}_1 + {\mathbf r}_2)/2$,
of a single dumbbell governed by Eq.~(\ref{eq:Langevin}) under the same force ${\mathbf f}$ acting on each bead
is
\begin{equation}
2 m_{\rm d} \ddot {\mathbf r}_{\rm cm}(t) = - 2 \gamma \dot {\mathbf r}_{\rm cm}(t) + 2 {\mathbf f}(t) + {\boldsymbol \xi}(t)
\label{eq:Langevin-single}
\end{equation}
with the new noise ${\boldsymbol \xi}(t) \equiv {\boldsymbol \eta}_1(t) + {\boldsymbol \eta}_2(t)$  with vanishing
average, $\langle \xi_a(t) \rangle =0$, and correlation 
\begin{equation}
\langle \xi_a(t) \xi_b(t') \rangle = 4\gamma k_BT \ \delta_{ab} \delta(t-t')
\; . 
\end{equation}
This is the Langevin equation of a point-like particle with mass $2m_{\rm d}$, under a force $2{\mathbf f}$ in contact with a bath with friction coefficient 
$2\gamma$ at temperature $T$. 
Equivalently, one can divide this equation by two and obtain a Langevin process for a point-like particle with mass 
$m_{\rm d}$ under a force ${\mathbf f}$ in contact with a bath with friction coefficient $\gamma$ at temperature $T/2$.
From these analogies one recovers several results on the statistics  of the center-of-mass position and 
velocity. Under no external force, ${\mathbf f}=0$, the center-of-mass velocity
is distributed according to the Maxwell distribution for a particle with mass $2m_{\rm d}$ at equilibrium at temperature $T$ (or mass $m_{\rm d}$ at temperature $T/2$).
The center-of-mass mean-square displacement between two times $t'$ and $t$
after preparation
%\begin{eqnarray}
\begin{equation}
\Delta^2(t,t') \equiv \langle [{\mathbf r}_{\rm cm}(t) - {\mathbf r}_{\rm cm}(t')]^2\rangle 
\end{equation}
%\nonumber\\
can be calculated as 
\begin{eqnarray}
\Delta^2(t,t') &=& d\Bigl[ \Bigl(v_{0}^{2}-\frac{k_BT}{2 m_{\rm d}}\Bigr) 
 \frac{(e^{-\frac{\gamma}{m_{\rm d}} t'} -e^{-\frac{\gamma}{m_{\rm d}}t})^{2}}{(\gamma/m_{\rm d})^{2}}
 \nonumber\\
&& \;\;\;+ 
 \frac{k_BT}{\gamma}(t-t') 
 \nonumber\\
 && \;\;\; -
\frac{m_{\rm d} k_BT}{\gamma^{2}}(1-e^{-\frac{\gamma}{m_{\rm d}}(t-t')}) \Bigl] .
\end{eqnarray}
where $v_0$ is the velocity of the particle at the initial time $t=0$.

Given the inertial time
\begin{equation}
t_I = m_{\rm d}/\gamma
\;  ,
\end{equation}
one obtains different limits in relation to different values of $t$ and $t'$.
At short times $0 \leq  t'\leq t \ll t_I$, by expanding all exponentials at small arguments, we obtain ballistic behaviour,
\begin{equation}
\Delta^2(t,t')=
 d v_0^2 \ (t-t')^2
\;  .
\end{equation}
At  long total times $t \geq t' \gg t_I$
and short time-delay 
$(t-t') \ll t_I$, 
one also obtains ballistic behaviour, 
\begin{equation}
\Delta^2(t,t')=
 d \ \frac{k_BT}{2m_{\rm d}} \ (t-t')^2
\;  ,
\end{equation}
with the initial velocity $v_0^2$ replaced by its average in equilibrium $\langle v^2\rangle = k_BT/(2m_{\rm d})$
(for a particle with mass $2m_{\rm d}$).
In both cases the time-delay dependence crosses over to diffusive motion
\begin{equation}
\Delta^2(t,t')=
2 d D^{\rm pm}_{\rm cm} \ (t-t')
\; , 
\end{equation}
for  long time delay $(t-t') \gg t_I$, with a diffusion constant taking the form
\begin{equation}
D^{\rm pd}_{\rm cm} = k_BT/(2\gamma)
\; . 
\label{eq:pddiffusion}
\end{equation}

This is the diffusion constant used in the definition 
of the P\'eclet number in Eq.~(\ref{eq:Peclet}).

The length of the dumbbell molecule, ${\mathbf R}(t) = {\mathbf r}_1 - {\mathbf r}_2$,
with ${\mathbf r}_1$ and ${\mathbf r}_2$ the position of the centers of the two spheres, is also 
a fluctuating quantity. For the parameters used one shows that $\langle R(t)\rangle$
approaches $R_\infty=0.96\approx\sigma_{\rm d}$.
The angular degrees of freedom can also be simply analyzed, especially under the assumption that $R(t)$ is constant
which is rather accurate since $R(t)$ does not fluctuate more than $3\%$ around its mean value.
In this approximation one finds that the angle diffuses with an angular diffusion constant equal to 
\begin{equation}
D_a=2k_BT/(\gamma R_\infty^2)
\; . 
\label{eq:DR}
\end{equation}

The {\it linear} instantaneous 
response, $R$, quantifies the effect of a small impulsive perturbation, say ${\mathbf h}(t'')$ applied at time 
$t''$, on the dynamics of the system of interest. 
For a dumbbell moving in a plane, 
 the $a$-th component of the center-of-mass position  perturbed by a force acting on its $b$-th component,
from time $t'$ until time $t$, is
\begin{equation}
\langle r_{\rm cm}^a(t) \rangle_{\mathbf h} = \langle r_{\rm cm}^a(t') \rangle + \int_{t'}^t dt'' \ R_{ab}(t,t'') \ h_b(t'')
\; . 
\label{eq:linear-response}
\end{equation}
Equivalently,
\begin{equation}
R_{ab}(t,t') \equiv \left. \frac{\delta \langle r_{\rm cm}^a(t) \rangle_{\mathbf h}}{\delta h_b(t')}  \right|_{{\mathbf h}=0} 
%= \frac{1}{2\gamma} \ \delta_{ab} \ \theta(t-t')
\; .
\end{equation}

Motivated by the interpretation of Eq.~(\ref{eq:Langevin-single})
that keeps the temperature of the noise unaltered and equal to $T$, we take the perturbation to the center-of-mass to
be ${\mathbf h}(t'')=2 {\mathbf f}(t'')$.
In order to  focus on the long time limit of interest, $t\gg t_I$, and to simplify the expressions, we 
drop the inertia term from Eq.~(\ref{eq:Langevin-single}). In this limit, the center-of-mass position is given by
\begin{equation}
{\mathbf r}_{\rm cm}(t) = {\mathbf r}_{\rm cm}(t') + \frac{1}{2\gamma} \int_{t'}^t dt'' \ [2{\mathbf f}(t'') + {\boldsymbol \xi}(t'')]
\; .
\label{eq:sol-com}
\end{equation}
The total linear response to a step-like perturbation that is applied from $t'$ to $t$,
with {\it independent} components $2f_a$ on each Cartesian spatial direction
such that $\delta (2f_a(t''))/\delta (2f_b(t')) =\delta_{ab}$, 
is then 
\begin{equation}
\chi(t,t') \equiv \int_{t'}^{t} dt'' \ R_{aa}(t,t'') = \frac{d}{2\gamma} \ (t-t')
\label{eq:chi-single}
\; . 
\end{equation} 
Summation over repeated indices was used to go from the second to third member of this equation.
Comparing now to the mean-square displacement,
$\Delta^2(t,t')$, one finds that in the diffusive regime 
\begin{equation}
2k_B T \chi(t,t') = \Delta^2(t,t')
\; , 
\end{equation}
and both functions depend on the two times only through their difference, $t-t'$.
This relation is the same as the one for a point-like Brownian particle in contact with a bath at 
temperature $T$.  We will use it as a reference to define the effective temperature from 
deviations in which the bath temperature is replaced by another 
parameter.
(The comparison between $\chi$ and 
the time-derivative of the correlation function between the position measured at different times,
 which in equilibrium is proportional to the inverse temperature of the 
bath,  yields a non-trivial relation 
that violates the equilibrium FDT for an unconfined Brownian particle~\cite{ckp}.)

A numerically convenient way 
to extract the linear response from Eq.~(\ref{eq:linear-response}), especially useful for the 
interacting systems studied in the following,
consists in taking the perturbing forces ${\mathbf f}$ to be uncorrelated with the center-of-mass position and random,
with zero mean, $[f_a]=0$, and correlation, $[f_a f_b ] = f^2 \delta_{ab}$.
Multiplying Eq.~(\ref{eq:linear-response}) by 
$2f_c$ and taking the average over their distribution one finds
\begin{equation}
[ \langle r_{\rm cm}^a(t) \rangle 2f_c] = (2f)^2 \int_{t'}^t dt'' R_{ac}(t,t'') 
\; . 
\end{equation} 
Setting $a=c$ and summing over components one obtains 
\begin{equation}
\chi(t,t') = \frac{[\langle r_{\rm cm}^a(t) \rangle 2f_a]}{(2f)^2} 
\; . 
\label{eq:chi-def-scalar-prod}
\end{equation}
Then,  one derives from Eq.~(\ref{eq:sol-com})
%In the single dumbbell case,  one derives from Eq.~(\ref{eq:sol-com})
\begin{equation}
[\langle r_{\rm cm}^a(t)\rangle 2f_a] = \frac{1}{2\gamma} \ \delta_{aa} \ (2f)^2 \ (t-t')
\; , 
\end{equation}
that yields
\begin{equation}
\chi(t,t') = \frac{d}{2\gamma} (t-t')
\end{equation}
consistently with Eq.~(\ref{eq:chi-single}).
%This method is useful especially in interacting systems.

The way in which we probe the linear response of the interacting system 
is by applying a force of modulus $2f$ in a random direction to the center-of-mass, 
and not on the rotational and vibrational degrees of freedom,
which gives
a linear response proportional to $1/(2\gamma)$ in the case of a single passive dumbbell. 
Averaging  over different angles,
% to minimise the numerical error,
one  recovers a result proportional to $1/(2\gamma)$. 
To obtain the linear response to  independent perturbations 
applied on the spatial $d$ directions, that is to say $\chi$,  we simply multiply the result by $d$.

\subsection{Passive many-body dumbbell system}

In order to study the many-body system we performed numerical 
simulations in the form described in Sec.~\ref{sec:algorithm}. We averaged over 100 different realizations of a system
with linear size $l=100$ 
and adimensional temperature $T=0.001$ (to make the notation 
lighter here and in what follows we avoid the asterisks to denote adimensional parameters). 
Measurements were done after 
an equilibration time of order $t_{\rm eq}=10^2$, starting from
an initial random configuration of our system, in both the position of the 
center-of-mass and the angular direction of the dumbbells,
and we show data gathered until a time equal to $10^6$.

\begin{figure}[h]
\begin{centering}
\includegraphics[scale=0.85]{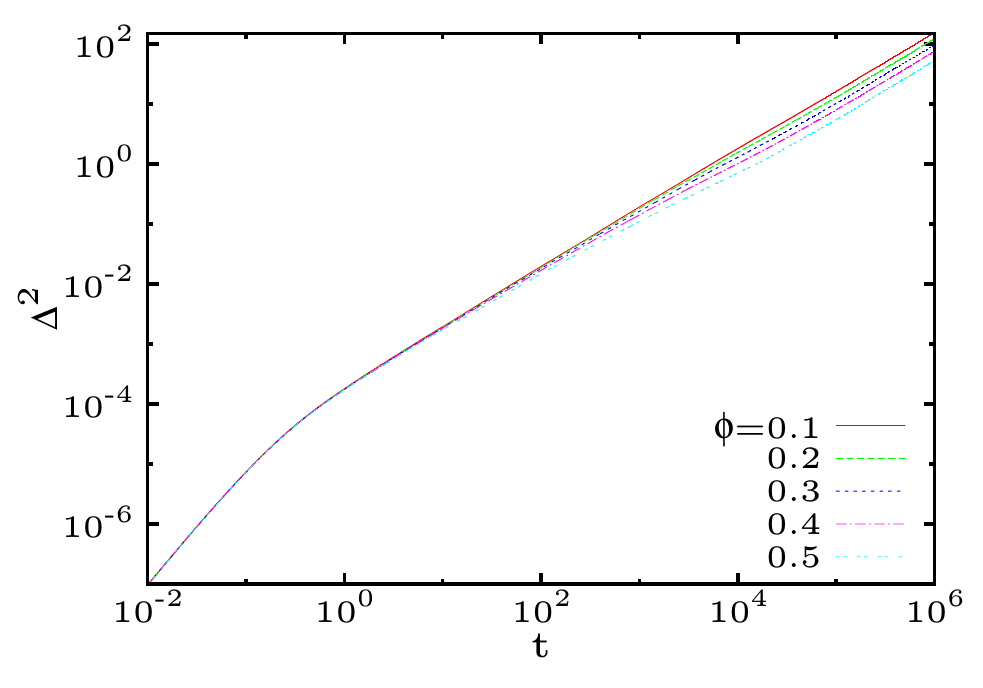}
\end{centering}
\caption{(Color online.) 
Mean-square displacement as a function of the delay time 
hereafter called $t$ measured after $t'=100$,
in a system of interacting passive dumbbells with the 
surface fractions given in the key at $T=0.001$ (double logarithmic scale). 
}
\label{fig:mean-square-displacement-passive-dumbbells}
\end{figure}

First, we studied a very loose system, with $\phi=0.01$ (data not shown). From the 
analysis of the system's global mean-square displacement  
we found that the ballistic regime ends at the inertial time $t_I = m_{\rm d}/\gamma \simeq
0.1$ when it crosses over to a diffusive regime. The ballistic regime is 
characterized by $\Delta^2(t) \simeq d \langle v^2 \rangle t^2$ with 
$\langle v^2 \rangle = k_BT/(2m_{\rm d})$ the thermal velocity of a Brownian particle 
with mass $2m_{\rm d}$, which for the parameters used takes the value 
$\langle v^2 \rangle \simeq 5 \cdot 10^{-4}$. The  diffusion constant in the free-diffusive regime is 
very close to the one of the single passive dumbbell, $D_{\rm cm} \simeq D_{\rm cm}^{\rm pd} 
\simeq 5 \cdot 10^{-5}$.

\begin{figure}[h]
\begin{centering}
\includegraphics[scale=0.9]{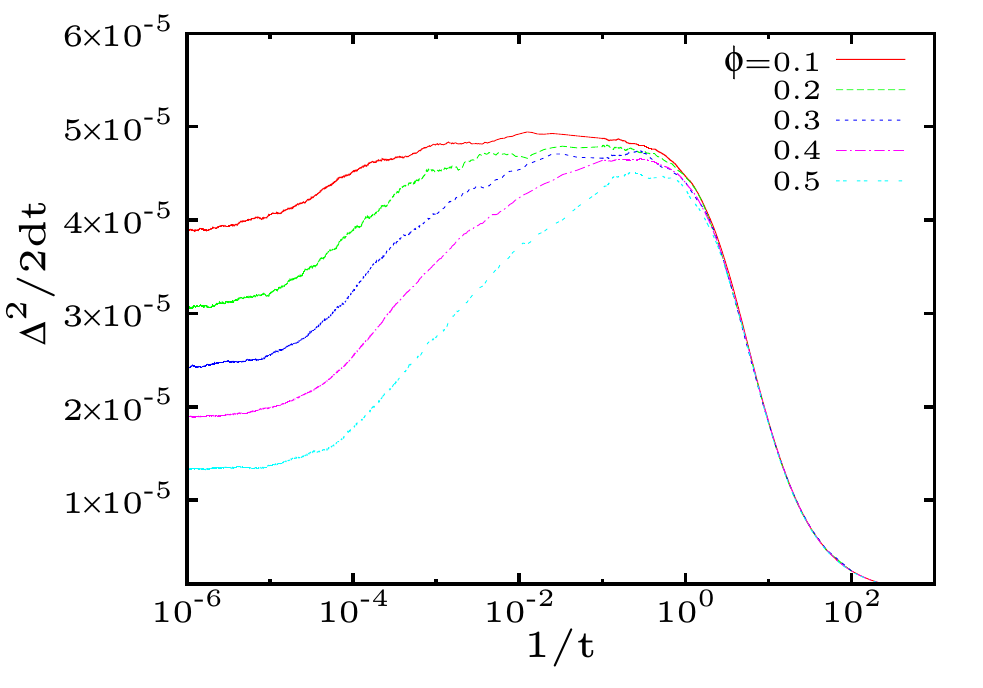}
\end{centering}
\caption{(Color online.) 
Rescaled mean-square displacement in systems of interacting passive dumbbells with the 
surface fractions given in the key. The data in Fig.~\ref{fig:mean-square-displacement-passive-dumbbells} 
are divided by $2dt$ and presented as a function of $1/t$ to identify the 
diffusive regime as a plateau with height given by the diffusion constant.
The single passive dumbbell diffusive constant takes the 
value $D_{\rm cm}^{\rm pd} = 5 \ 10^{-5}$ for these parameters, consistently with the trend of the 
numerical results for the many-particle system.
}
\label{fig:zoom-over-ballistic-diffusive-passive-dumbbells}
\end{figure}

Next,  we studied systems with five higher surface fractions: 
$\phi=0.1, \ 0.2, \ 0.3, \ 0.4, \ 0.5$, see Fig.~\ref{fig:mean-square-displacement-passive-dumbbells}. 
We found that, as in the single molecule case, the ballistic regime crosses over at $t_I = m_{\rm d}/\gamma =
0.1$  to a first free diffusive regime, approximatively in the interval $1 \leq t \lesssim 10^2$,  that is now followed by 
a second diffusive regime that feels the effect of the dumbbell concentration~\cite{dhont}.
Quite naturally, the dynamics slows down for denser systems as shown by the spread of curves at the late stages
in the plot.

\begin{figure}
\begin{center}
  \begin{tabular}{cc}
      \includegraphics[scale=0.9]{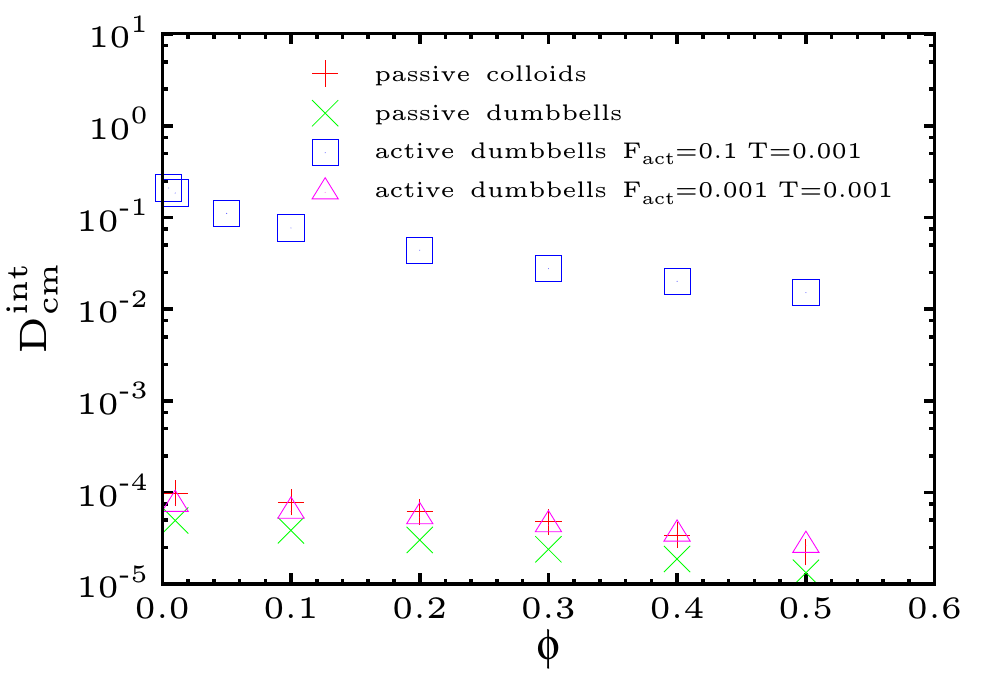}\\%(a)\\
    \includegraphics[scale=0.9]{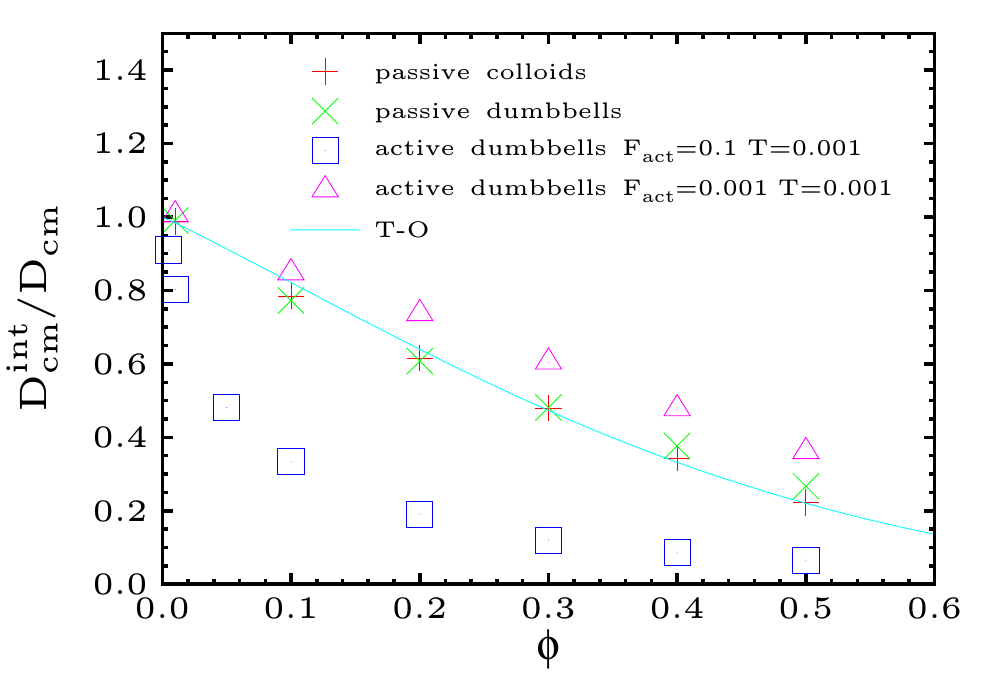} %\\(b)\\
    \end{tabular}
\caption{(Color online.) 
Upper panel: in linear-log scale, the diffusion constant of an interacting system, as a function of 
surface fraction $\phi$, in four cases:
the passive dumbbell system displayed with (green) crosses $\times$;
the passive colloidal system shown with (red) pluses $+$;
the active dumbbell system with $F_{\rm act}=0.001$ and $F_{\rm act}=0.1$ drawn with 
(pink) triangles $\triangle$ and 
(blue) squares $\square$, respectively. 
In all cases the systems are in contact with the same thermal bath at temperature $T=0.001$
and interact with the same WCA potential given in Eq.~(\ref{eq:WCA-potential}).
Lower panel: in double linear scale, the same data
relative to its single particle limit, as a function of 
surface fraction $\phi$.
The solid line represents the analytic prediction by Tokuyama-Oppenheim (TO) for a colloidal system with the 
same parameters~\cite{Tokuyama}. 
}
\label{fig:diffusive-constant-passive-dumbbells} 
\end{center}
\end{figure}

In Fig.~\ref{fig:zoom-over-ballistic-diffusive-passive-dumbbells}
data are represented in such a way that the last diffusive regime
appears as a plateau  at $1/t\to 0$ with height equal to the diffusive constant.
The diffusive constant in the interacting regime, normalized by the one 
of the center-of-mass of the free passive dumbbell, $D_{\rm cm}^{\rm pd}$, is plotted in
Fig.~\ref{fig:diffusive-constant-passive-dumbbells} against the surface fraction. For comparison we also similarly display  normalised
data obtained for a system of colloids coupled to the same thermal bath and interacting via 
the WCA potential~(\ref{eq:WCA-potential}). Within numerical accuracy the colloidal and dumbbell systems 
have the same behaviour  until $\phi \lesssim 0.3$ while the dumbbell data are 
slightly above the colloidal ones for larger values of $\phi$. 
We find very good agreement with the prediction of Tokuyama and Oppenheim,
\begin{eqnarray}
\frac{D_{\rm cm}(\phi)}{D_{\rm cm}(0)}
=\frac{1}{1+H(\phi)}
\end{eqnarray}
with 
\begin{eqnarray}
H(\phi) &=& 
\frac{2[b(\phi)]^2}{1-b(\phi)}
-\frac{c(\phi)}{1+2c(\phi)}
\nonumber\\
&& -\frac{b(\phi) \ c(\phi) \ [2+c(\phi)]}{[1+c(\phi)] \ [1-b(\phi)+c(\phi)]} 
\end{eqnarray}
and 
\begin{eqnarray}
b(\phi) = \sqrt{9\phi/8}   \qquad c(\phi) =11\phi/16
\; , 
\end{eqnarray}
derived from a perturbative calculation 
for low-concentrated colloidal systems~\cite{Tokuyama}. This  expression has no free parameters. 
The agreement  is very good for all  surface fractions 
for colloids, and until $\phi \simeq 0.3$ for dumbbells (continuous line in the figure)
if we use the three terms above in the expansion.
For denser systems  the shape of the molecules starts playing a role. 
We stress that the comparison is made 
between normalized curves. 
The third and fourth sets of data in this figure correspond to systems of active dumbbells
and will be discussed in Sec.~\ref{sec:active-many-body}. 

\section{Active single molecule}
\label{sec:single-active}

We turn now to the dynamics of a single active  
molecule as a reference case for the  interacting system to be discussed in the next section.

Take a single active molecule constrained to move in two dimensional space.
The length of the molecule is not altered by the active force and it still approaches $R_\infty\approx\sigma_d$.
From the Langevin equation for the center-of-mass position that acquires 
a forcing term one calculates the average square velocity of the center-of-mass.
For $t\gg t_I$ the stationary value for its $a$-th component is given by
\begin{eqnarray}
\langle {v^2_{\rm cm}}_{a}(t)\rangle 
%&=& \frac{1}{2m_{\rm d}}\bigg[k_BT+\frac{F_{\rm act}^2}{\gamma(\frac{\gamma}{m_{\rm d}}+D_R)}\bigg]
%\nonumber\\ 
&\to & \frac{1}{2m_{\rm d}}\bigg(k_BT+\frac{F_{\rm act}^2}{\gamma(\frac{1}{t_I}+\frac{1}{t_a})}\bigg)
\nonumber\\
&\equiv& \frac{k_B T_{\rm kin}}{2m_{\rm d}}
\; , 
\label{kinetic_temperature_formula}
\end{eqnarray}
with the inertial and angular time-scales
 \begin{equation}
 t_I = \frac{m_{\rm d}}{\gamma}
 \; , 
 \qquad
 \qquad
 t_a = \frac{1}{D_a} =  \frac{\gamma\sigma_{\rm d}^2}{2k_BT}
 \; , 
 \label{eq:time-scales}
 \end{equation}
 that will find a clear meaning when studying the mean-square displacement below.

The formula (\ref{kinetic_temperature_formula}) gives an expression to what is called the kinetic or granular 
temperature~\cite{Jaeger96,Aronson06,Pouliquen08} of
the system.  Quite naturally, $T_{\rm kin}=T$ for $F_{\rm act}=0$. 
Note that $T_{\rm kin}$ depends on the two time scales $t_I$ and $t_a$. It approaches 
\begin{eqnarray}
k_B T_{\rm kin} &\simeq& 
\left\{
\begin{array}{ll}
k_BT + t_I F_{\rm act}^2/\gamma 
\;\;\;\;\;\;\;\; & \mbox{for} \;\;\;\;\;\;\;\; t_a \gg t_I
\; ,
\nonumber\\
k_BT + t_a F_{\rm act}^2/\gamma
\;\;\;\;\;\;\;\; & \mbox{for} \;\;\;\;\;\;\;\; t_I \gg t_a
\; . 
\end{array}
\right.
\end{eqnarray}
In our problem $t_a\gg t_I$ and the relevant limit is 
\begin{equation}
k_B T_{\rm kin} = k_B T + \frac{m_{\rm d}}{\gamma^2} F_{\rm act}^2
\; . 
\end{equation}
Moreover, for the parameter values later used in the simulation, $m_{\rm d}/\gamma^2 = 0.01$.

\begin{figure}[h]
\includegraphics[scale=0.9]{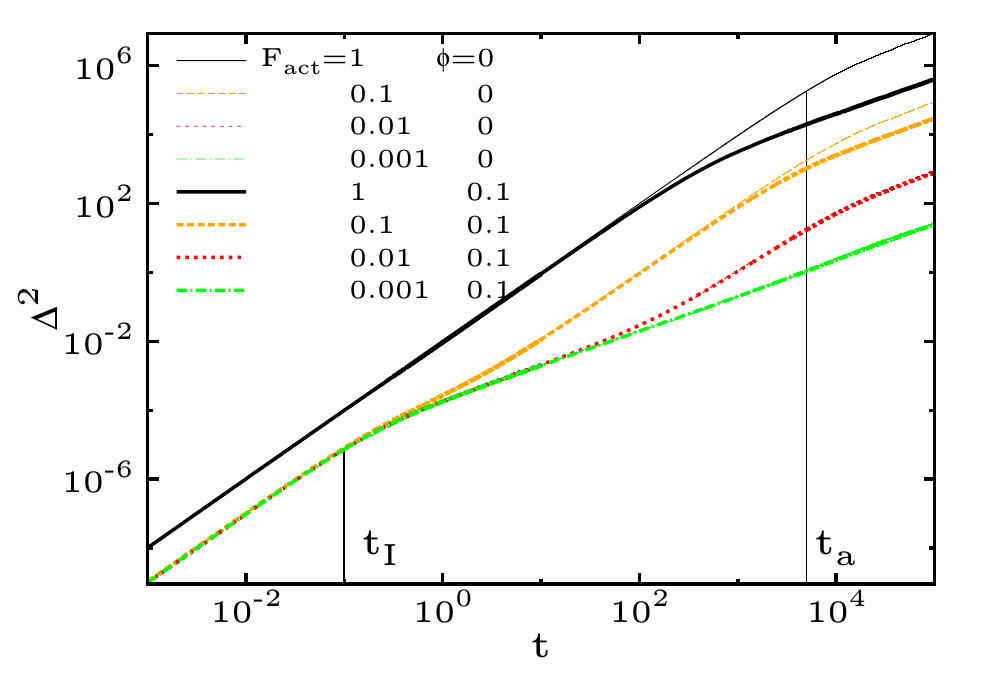}
\caption{(Color online.) 
Mean-square displacement of active dumbbells
for various values of $F_{\rm act}$ given in the key
at temperature $T=0.001$ and linear size of the system  $l=100$. The behaviour of a single dumbbell is shown with 
thinner
lines while the one of a system of interacting dumbbells with $\phi=0.1$ is shown with thicker lines.  
The average has been taken over 
$200$ thermal histories.
The inertial and angular time-scales are signaled with vertical black lines at $t_{I}=0.1$ and $t_a=5000$, respectively.
The floating time-scale $t^*$ is proportional to the inverse square active force, $F_{\rm act}^{-2}$. See the 
text for a discussion of its effect. 
}
\label{fig:dumbattivevarifi_1}
\end{figure}

After some simple calculations  one finds  that 
the mean-square displacement of a single dumbbell molecule moving on a plane under the longitudinal active force 
has an initial ballistic regime for times shorter than the 
inertial time-scale $t_I$. Beyond this time-scale, 
the over-damped limit takes over and inertia can be neglected. One 
can then quite easily derive the time-behaviour of the self-diffusion. 
For times shorter than the angular time scale $t_a$ one finds 
\begin{equation}
\Delta^2(t) \simeq 2d D^{\rm pd}_{\rm cm} \ t +  \left( \frac{F_{\rm act}}{\gamma} \right)^2 t^2
\end{equation}
with $D^{\rm pd}_{\rm cm}=k_BT/(2\gamma)$ the diffusion constant of the passive diatomic molecule (see also~\cite{Lowen11} 
where a similar calculation for an active ellipsoid has been performed).
Within this regime one can still identify two sub-regimes. For times shorter than 
$t^* = 2 d D^{\rm pd}_{\rm cm} \gamma^2/F^2_{\rm act}$ the first term dominates and the 
dynamics is diffusive as in the absence of the active force. If, instead, times are
longer than $t^*$ the dynamics becomes ballistic again and it is controlled by the 
active force. For even longer times, going beyond $t_a$, a new diffusive 
regime establishes, 
\begin{equation}
\Delta^2(t) \simeq 2dD_A \ t
\; , 
\end{equation}
 with a diffusion constant that depends upon the 
active force and in $d=2$ is equal to
\begin{eqnarray}
D_A &=& D^{\rm pd}_{\rm cm} + \frac{1}{2} \left( \frac{F_{\rm act}}{\gamma} \right)^2 \frac{1}{D_a}
.
\end{eqnarray}
In order to make the parameter dependence explicit we call this 
diffusion constant
\begin{eqnarray}
\!\!\!\! D_A(F_{\rm act}, \phi=0)
\! &=& \! \frac{k_BT}{2\gamma} \! \left[ 1 + \frac{1}{2}  \! \left(\frac{ F_{\rm act} \sigma_{\rm d}}{k_BT}\right)^2 \right]
\label{active_diffusion_constant}
\end{eqnarray}
with $R_\infty$ replaced by $\sigma_{\rm d}$. 
This expression shows a dependence on the square 
of the P\'eclet number that will later appear in the effective temperature as well.
The reason why the dynamics slow down in this last regime although there 
is still the active force acting on the dumbbell is that the angular motion goes against
the translational one. All these regimes are summarized in
 \begin{eqnarray}
&&
 \mbox{ballistic} \mapsto \mbox{diffusive} \mapsto \mbox{ballistic} \mapsto \mbox{diffusive}
 \nonumber
 \\
&& \qquad\;\;\;\;\; t_I \qquad\qquad\;\; t^* \qquad\;\;\;\;\;\; t_a
\nonumber
 \end{eqnarray}
 with the time-scales $t_I$ and $t_a$ introduced in Eq.~(\ref{eq:time-scales}).
% \begin{equation}
% t_I = \frac{m_{\rm d}}{\gamma}
% \qquad
% \qquad
% t_a = \frac{\gamma\sigma_{\rm d}^2}{2k_BT}
% \; . 
 %\end{equation}
 Note that the ``intermediate" time-scale $t^*$ is inversely proportional to the square of the active force, $F^{-2}_{\rm act}$, and can
 go below the inertial time-scale $t_I$ for sufficiently strong non-equilibrium forcing or beyond $t_a$ for sufficiently weak one.

The generic features described in the previous paragraph
can be seen in Fig.~\ref{fig:dumbattivevarifi_1},  
especially in the curves for $F_{\rm act}=0.1, \ 0.01$, where $t^* < t_a$ and the 
 regimes separated by $t^*$ are distinct.  
 $t^*$ goes below $t_I$  for the largest active force $F_{\rm act}=1$ used in Fig.~\ref{fig:dumbattivevarifi_1}
for which the data shown have already reached 
the second ballistic regime and cross over, at $t_a$, to the last diffusive regime. At the other
extreme, for the smallest active force $F_{\rm act}=0.001$ the time-scale $t^*$ is very large
going beyond the time-scale $t_a$. The data show the first ballistic regime, the cross-over to the 
first free-diffusive regime and a very smooth cross-over to the last diffusive regime with a 
different diffusion constant. There is no second ballistic regime in this case.
For the intermediate active forces, $F_{\rm act}=0.1, \ 0.01$, the four regimes can be 
observed in the data curves. 
 The numerical results yield values for the cross-over times, diffusion constants and 
 velocities in the ballistic regimes that are indistinguishable -- within numerical 
 accuracy -- from the analytic predictions. 

A similar crossover in the mean-square displacement was 
observed in systems and models of spherical particles. On the theoretical side, 
reference to such a crossover is made in~\cite{Fily12,Levis-Berthier,Lowen11}. 
In~\cite{Howse07} the dynamics of artificial swimmers made of polystyrene 
microspheres coated with platinum on one size, while keeping
the second half as the non-conducting polystyrene, was studied. In this 
system,  the platinum catalyzes the reduction of a `fuel' of hydrogen peroxide 
to oxygen and water propelling the particles in a preferred direction.
Particle tracking was used to characterise the particle motion 
as a function of hydrogen peroxide concentration. 
At short times, the particles move predominantly in a directed way, 
with a velocity that depends on the concentration of the fuel molecules,
while at long times the motion randomises and becomes diffusive, with a 
diffusion constant that also depends on the fuel concentration. The 
analytic results for the single dumbbell mean-square displacement are 
consistent with the experimental results. A similar crossover was also
observed  in the motion of beads propelled by adsorbed bacteria~\cite{Darnton04}
and in bacterial baths~\cite{wu}.

Another, independent probe of the dynamics of an out of equilibrium system is the  
displacement induced by applying a small constant external force to one (or more)  tagged
particle(s) in the system:
\begin{eqnarray} 
\chi(t) & \equiv & \frac{d}{N (2f)^2} \sum_{i=1}^N \langle 2{\mathbf f}_i \cdot {\mathbf r}_{{\rm cm}_i}(t) \rangle 
\nonumber\\
&=& \frac{d}{N (2f)^2} \sum_{i=1,3,5,...2N} \langle 2{\mathbf f}_i \cdot \frac{{\mathbf r}_{i}(t)+{\mathbf r}_{i+1}(t)  }{2} \rangle 
\nonumber\\
&=& \frac{d}{N (2f)^2} \sum_{i=1}^{2N} \langle {\mathbf f}_i \cdot {\mathbf r}_{i}(t) \rangle
\; . 
\label{eq:def-chi}
\end{eqnarray}
This equation generalises (\ref{eq:chi-def-scalar-prod}) to the many-body system.

The perturbing force ${\mathbf f}_i ={\boldsymbol \epsilon}_i f$ is applied to every monomer
 (in the same way as the active force) at time $t_0$ and kept  constant until
the measuring time $t$.
$f$ is its modulus and ${\boldsymbol \epsilon}_i$ its direction which is uniformly distributed
and is the same for the two monomers of a given dumbbell.
In the case of a single active dumbbell $N=1$ and  one finds again
\begin{equation}
\chi(t) = d (2 \gamma)^{-1} \ t
\label{eq:chi-single-dumbbell}
\end{equation}
a result that is independent of the activation force $F_{\rm act}$. 
We will write it as
\begin{equation}
\chi(t) = d\mu(F_{\rm act},\phi=0) \ t
\label{eq:chi-single-dumbbell}
\end{equation}
with $\mu(F_{\rm act},\phi=0)=\mu(F_{\rm act}=0,\phi=0)$.
Using the 
relation between mean-square displacement, $\Delta^2$, and  induced displacement, 
$\chi$, in equilibrium, $2 k_B T \chi(t) = \Delta^2(t)$  for all $t$, to define a possibly time-dependent effective 
temperature out of equilibrium
\begin{equation}
2 k_B T_{\rm eff}(t) \chi(t) = \Delta^2(t) 
\; , 
\end{equation} 
one finds, in the late diffusive regime $t\gg t_a$, 
\begin{eqnarray}
k_BT_{\rm eff}(t) &\to& 
k_BT_{\rm eff}(F_{\rm act}, \phi=0) 
\nonumber\\
&=&  \frac{D_A(F_{\rm act}, \phi=0)}{\mu(F_{\rm act}, \phi=0)}
\end{eqnarray}
that becomes
\begin{eqnarray} 
 T_{\rm eff}(F_{\rm act}, \phi=0) 
 &=&  T \left[1 + \frac{1}{2} \left(\frac{F_{\rm act}\sigma_{\rm d} }{k_BT}\right)^2\right] 
\nonumber\\
&= &
 T \left( 1+ \frac{1}{8} \mbox{Pe}^2 \right)
 \; . 
 \label{T_eff_equation}
\end{eqnarray}
%for $d=2, \ 3$.
We can arrive at the same result by defining  $T_{\rm eff}$ from the Einstein relation between diffusion constant and mobility.
We stress the fact that $T_{\rm eff}$ takes a constant value in this late diffusive regime. Note the 
different bath-temperature dependence in $T_{\rm eff}$ and $T_{\rm kin}$.
We reckon also that $T_{\rm eff} \geq T_{\rm kin}$ and that the two expressions coincide in the (unphysical) case $t_I\gg t_a$.
For the parameter values later used in the simulations $k_BT_{\rm eff} = k_BT + 0.5 \ F_{\rm act}^2/(k_BT)$
and the factor in the $F_{\rm act}^2$ term in $k_BT_{\rm eff}$ equals 10 at $T=0.05$ (while the one in $k_BT_{\rm kin}$ is only 0.01).

The response-displacement relation in a stochastic model for an active particle was studied in~\cite{Ben-Isaac-etal}.
In this paper, an overdamped Langevin equation for a randomly kicked point-like particle with a variable number of kicking motors
producing  pulses of force $\pm f_0$ acting at Poisson distributed times during intervals of duration $\Delta \tau$, and in contact with a thermal environment,
was analyzed.
% in~\cite{Ben-Isaac-etal}. 
In the low frequency limit (long time-delays) the effective temperature approaches a constant 
$T_{\rm eff} \to T + N_m f_0^2 \Delta\tau^2/(\tau + \Delta \tau)$ with $N_m$ the number of motors and $\tau$ the averaged waiting-time 
between the pulses. The dependence on $f_0^2$ is similar to the dependence on Pe$^2$ that we find for the single active dumbbell.

Again from the $T_{\rm eff}$ perspective, 
Szamel~\cite{Szamel14} studied a different model in which the self-propulsion of a single point-like active particle is modelled as a fluctuating 
force evolving according to the Ornstein-Uhlenbeck process, independently of the state of the particle. The particle moves in a viscous medium 
that is assumed to force overdamped motion. The free diffusion properties of the particle lead to an effective temperature 
defined from an extension of the Einstein relation linking the diffusion coefficient of the free particle to the variance of the fluctuating term in the
Ornstein-Uhlenbeck process for the active force and the friction coefficient of the medium. 
The mobility and self-diffusion of an active Janus 
particle were monitored with micro-tracking by Palacci {\it et al.}~\cite{Palacci10} to infer from them the effective temperature. 
In both cases, as for our active single 
dumbbell, the effective temperature increases with the activation parameter as Pe$^2$. 

Palacci {\it et al.}~\cite{Palacci10}, Tailleur and Cates~\cite{Tailleur08,Tailleur09} and Szamel~\cite{Szamel14}
measured and calculated the stationary probability distribution of the active particles' positions
in a linear external potential (mimicking gravity) in a regime such that the sedimentation velocity is small with respect
to the swimming velocity. They
found an exponential form that corresponds to a Boltzmann distribution under gravity, with the parameter associated to the sedimentation length 
given by the equilibrium one with the temperature being replaced by the effective one of the free active particle.
 Szamel also studied whether 
the ambient temperature is simply replaced by the single particle effective one 
in the position probability distribution function of the Ornstein-Uhlenbeck
active particle under a different external potential and found that this is not the case~\cite{Szamel14}.

\section{Active many-body dumbbell system}
\label{sec:active-many-body}

The collective behavior of an ensemble of dumbbell molecules in interaction and under
the effect of active forces is very rich~\cite{Suma13, Suma14}. 
In this section we first describe some general features of this behavior
in the homogeneous phase. Later, we will focus on the evaluation of the diffusion constant,
linear response  and effective temperature. 

\subsection{Homogeneous phase}

\begin{figure}[h]
\begin{centering}
\includegraphics[scale=0.9]{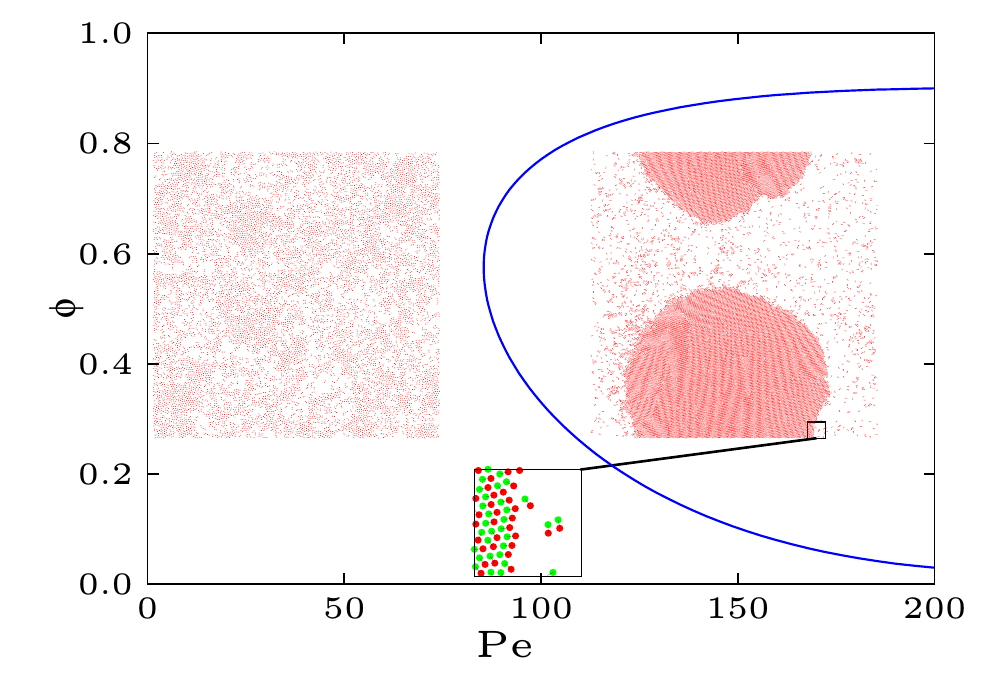}
\end{centering}
\caption{(Color online.) Schematic phase diagram of the dumbbell system. 
Inside the curve, at high P\'eclet numbers, the system undergoes 
phase separation into two phases characterized by  two different densities.  Here, 
typically, large  and stable clusters are observed, as shown in the snapshot on the right. The dumbbells are freezed and point preferentially
towards  the center of the cluster. The enlargement shows the border of such a cluster, with  the green part, 
corresponding to the head of the dumbbells, pointing inside,  and the tails (red part) pointing outside. For small P\'eclet numbers
 the system does not show the formation of such large clusters and a typical configuration is presented.
The location of the transition line is based on the results of simulations with $F_{\rm act}=1$ (see \cite{Suma13, Suma14} for further details);
the right snapshot is taken at $T=0.01$, the one on the left at $T=0.05$.
}
\label{fig:phasediagram}
\end{figure}

\begin{figure}[b]
\begin{center}
     \includegraphics[scale=0.37]{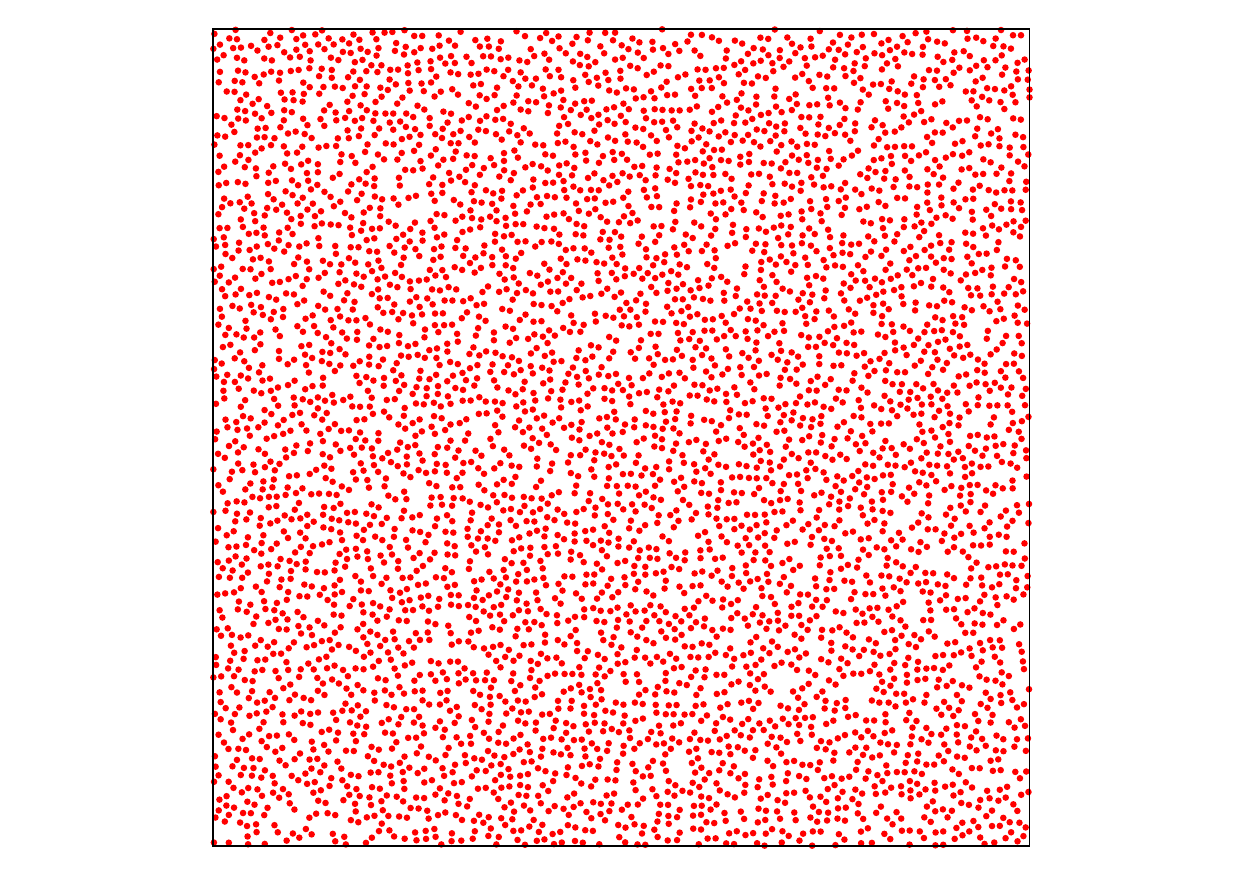}
     \hspace{-1cm}
     \includegraphics[scale=0.37]{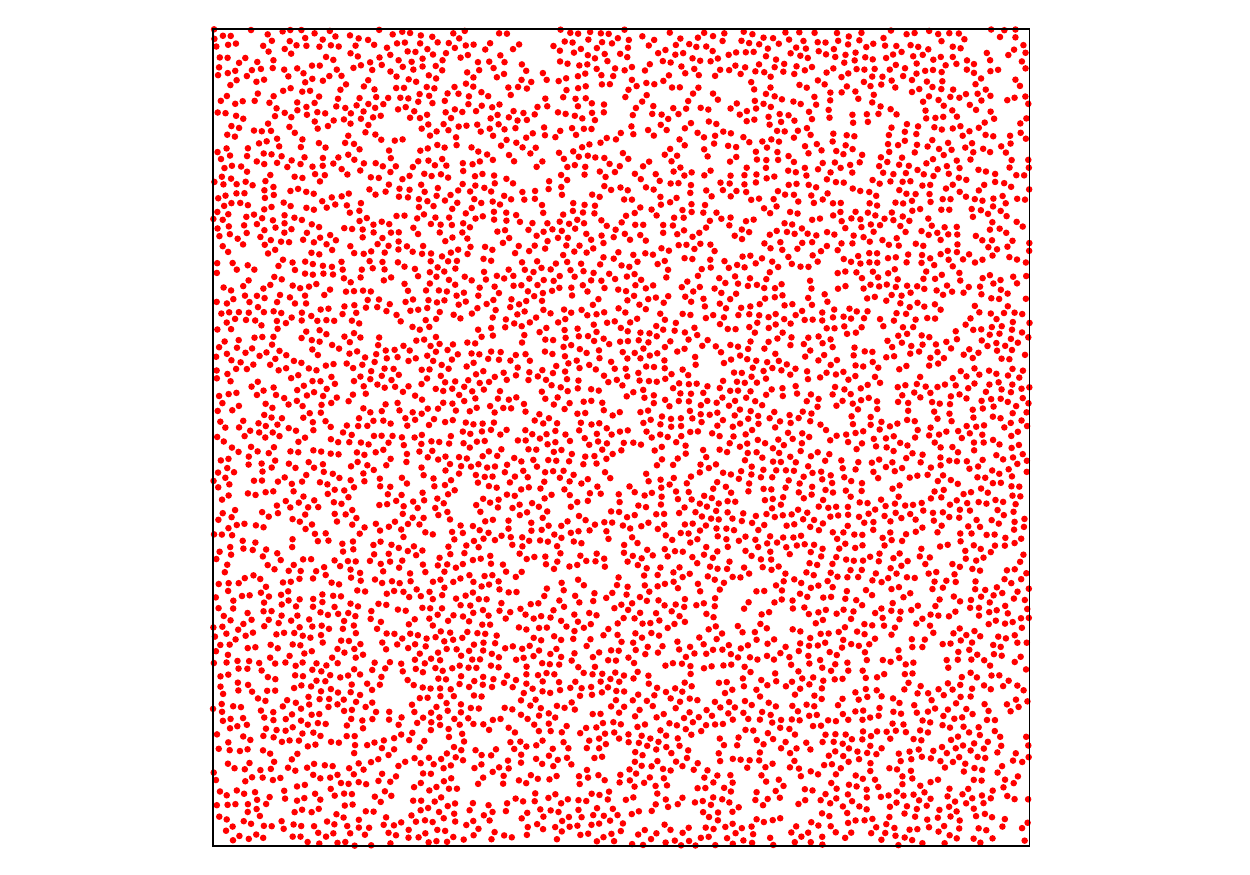}
         \\   %  (a)   \hspace{3.4cm}      (b)\\
 \end{center}
 \begin{center}
     \includegraphics[scale=0.37]{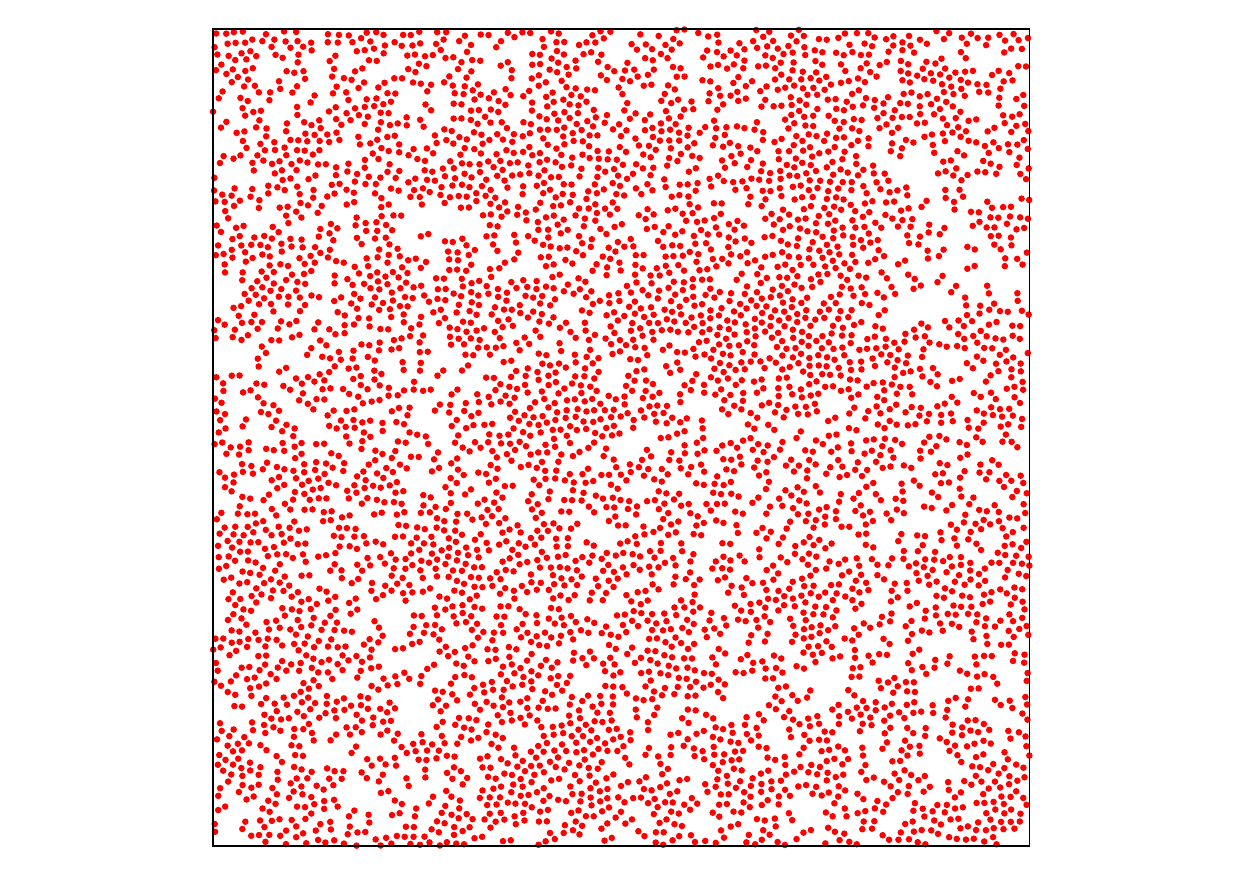}
        \hspace{-1cm}
     \includegraphics[scale=0.37]{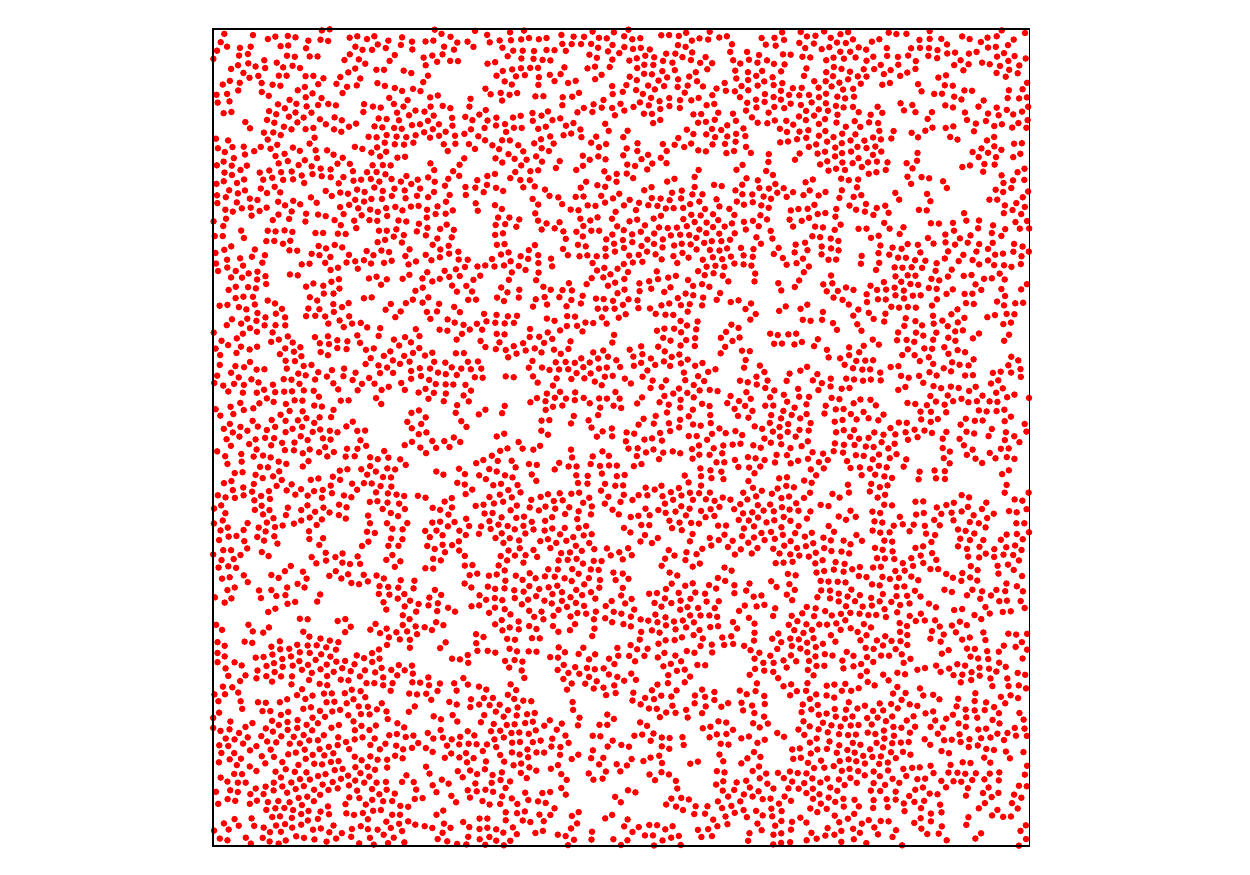}
          \\  %   (c)   \hspace{3.4cm}      (d)\\
 \end{center}
 \begin{center}
     \includegraphics[scale=0.37]{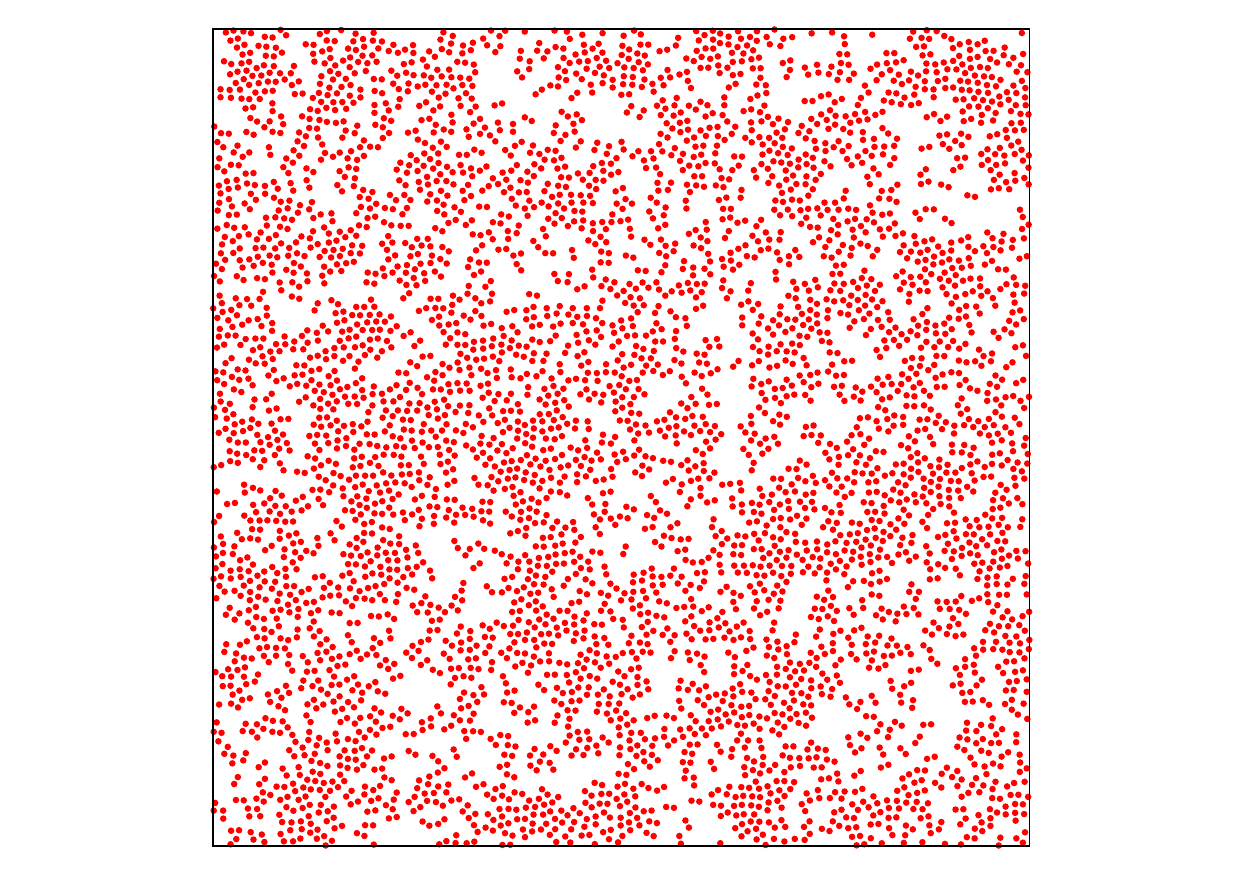}
        \hspace{-1cm}
     \includegraphics[scale=0.37]{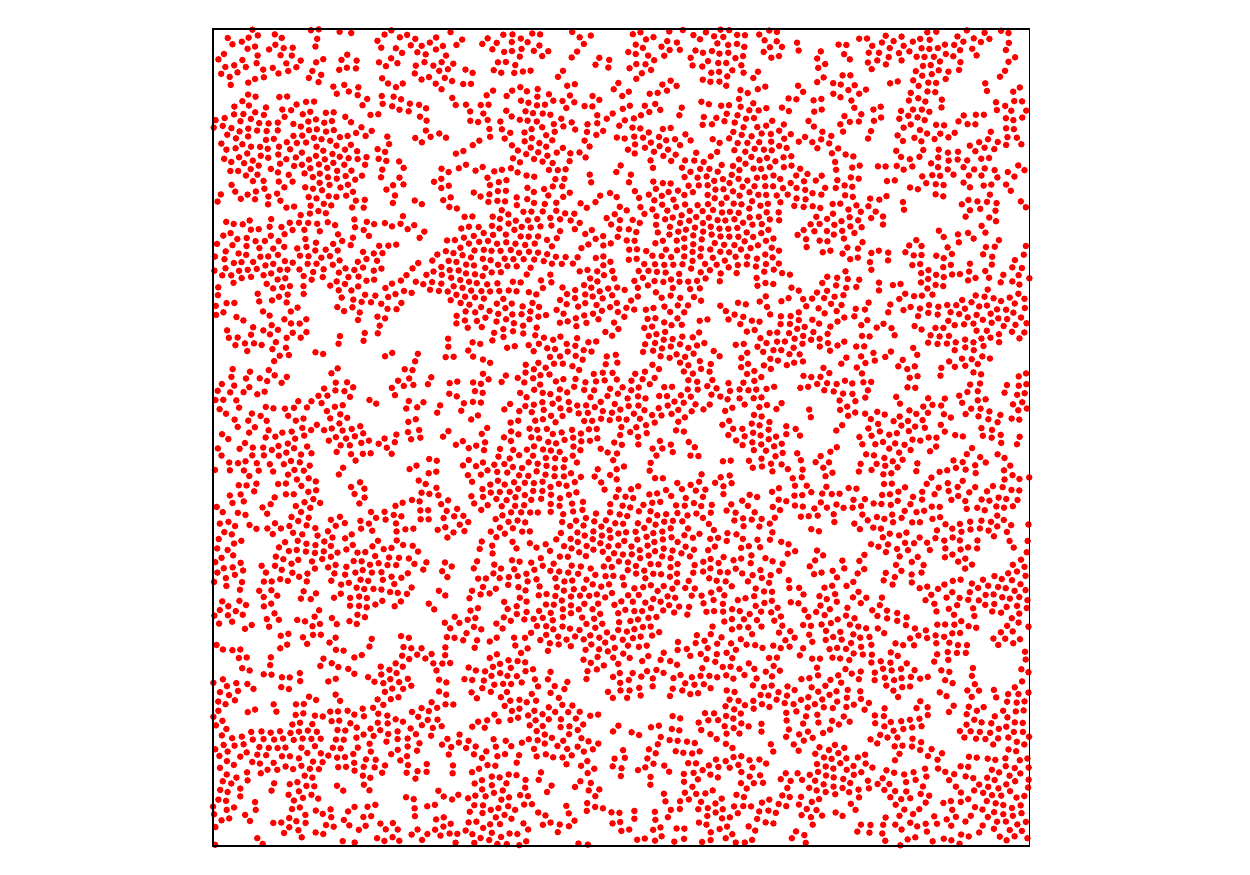}
          \\   %  (e)   \hspace{3.4cm}      (f)\\
 \end{center}
%    \\
%   \includegraphics[scale=0.34]{FIGS/vxdumbfi0_3}
\caption{(Color online.) 
Clustering effects in the active many-body system. Density is $\phi=0.4$, temperature is $T=0.05$ and 
from left to right and top to bottom the active forces are $F_{\rm act}= 0.01, \ 0.1, \ 0.3, \ 0.5, \ 0.7, \ 1$.
The side of the box is $l=100$.
}
    \label{fig:clusters}
\end{figure}

Active dumbbell systems exhibit  a transition between a 
%homogeneous
phase at small P\'eclet number,  that includes the 
%similar to the one of the
 passive limit, 
and a phase with stable aggregates of dumbbells, not existing without activity,
%clustering and large density fluctuations 
at high P\'eclet number. 
A schematic representation of the phase diagram is given in Fig.~\ref{fig:phasediagram}.
We will not discuss the details of 
this
%non-equilibrium 
phase diagram here but we simply state that we will work with sufficiently
low P\'eclet number so as to stay in the region that we call
%within
 the homogeneous phase 
although clustering phenomena, that will be important for the 
effective temperature, are already present. 
Typical configurations for the two phases are also shown in Fig.~\ref{fig:phasediagram}. 
In the aggregated phase the clusters reach the size of the system, while 
in the homogeneous phase they remain of finite size and are not stable.
The phase separating kinetics in the  phase with large-scale aggregation  was studied in~\cite{Suma13}.
One can also observe that, while in the  clusters of the homogeneous phase
the dumbbells are not stucked and can move, they are frozen inside  the aggregates in  the high P\'eclet 
 phase.

\begin{figure}[t]
\begin{center}
     \includegraphics[scale=0.37]{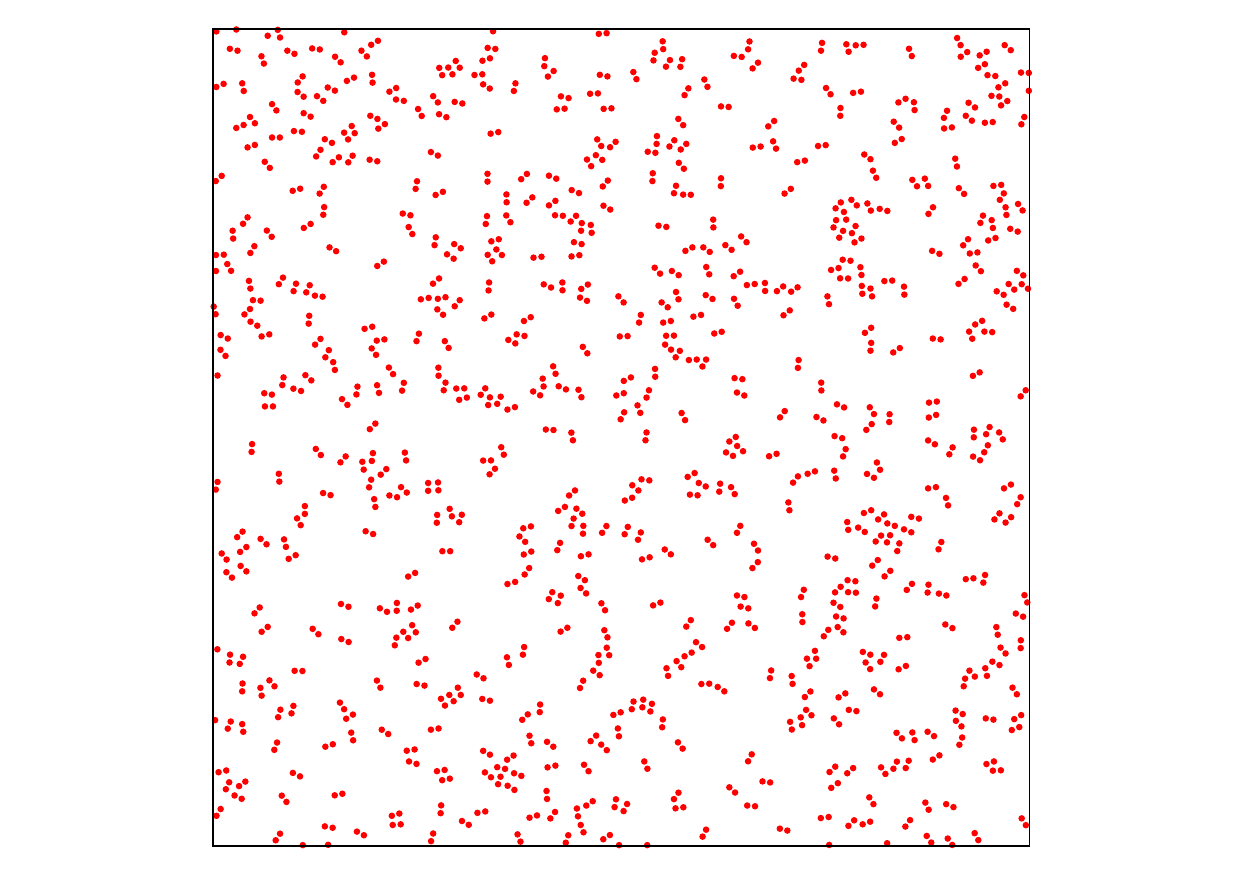}
     \hspace{-1cm}
     \includegraphics[scale=0.37]{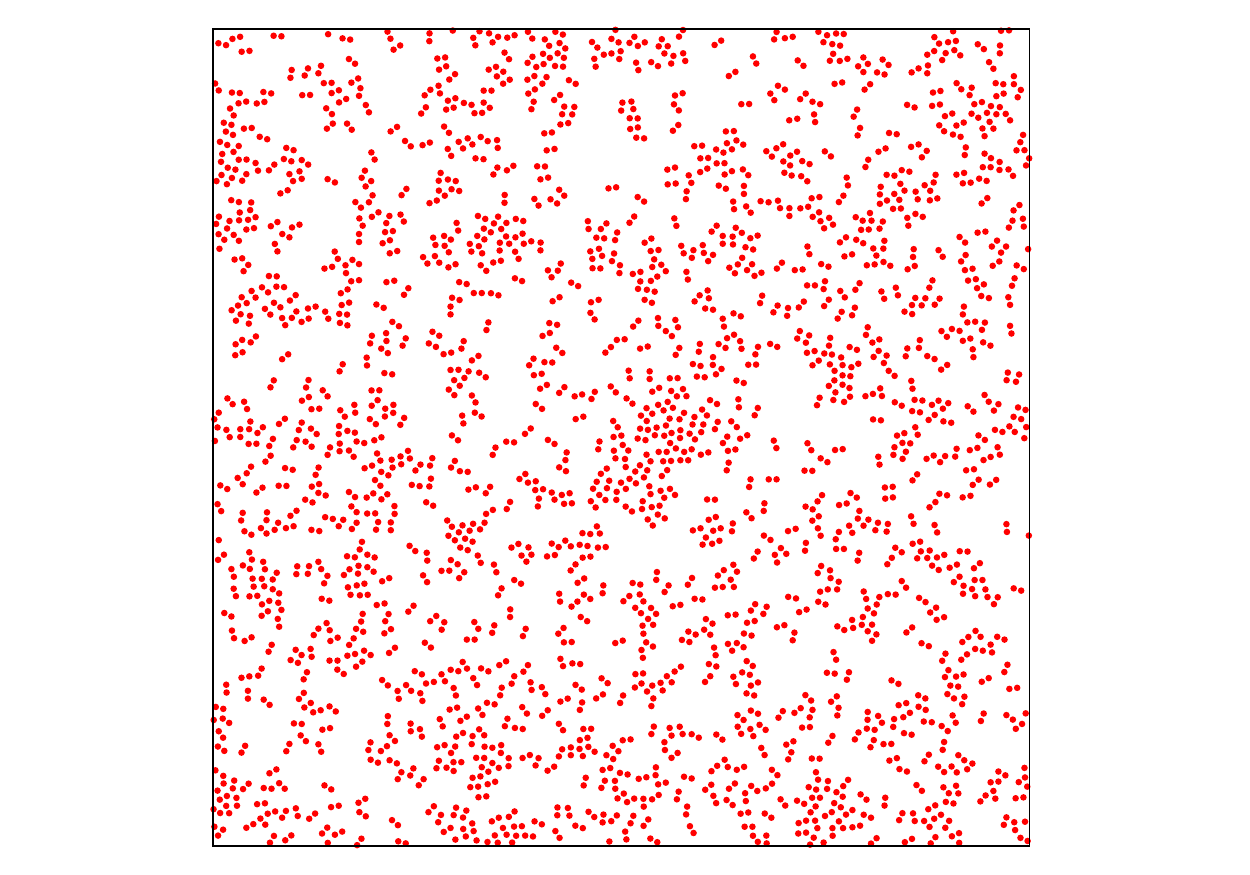}
          \\    % (a)   \hspace{3.4cm}      (b)\\
 \end{center}
 \begin{center}
     \includegraphics[scale=0.37]{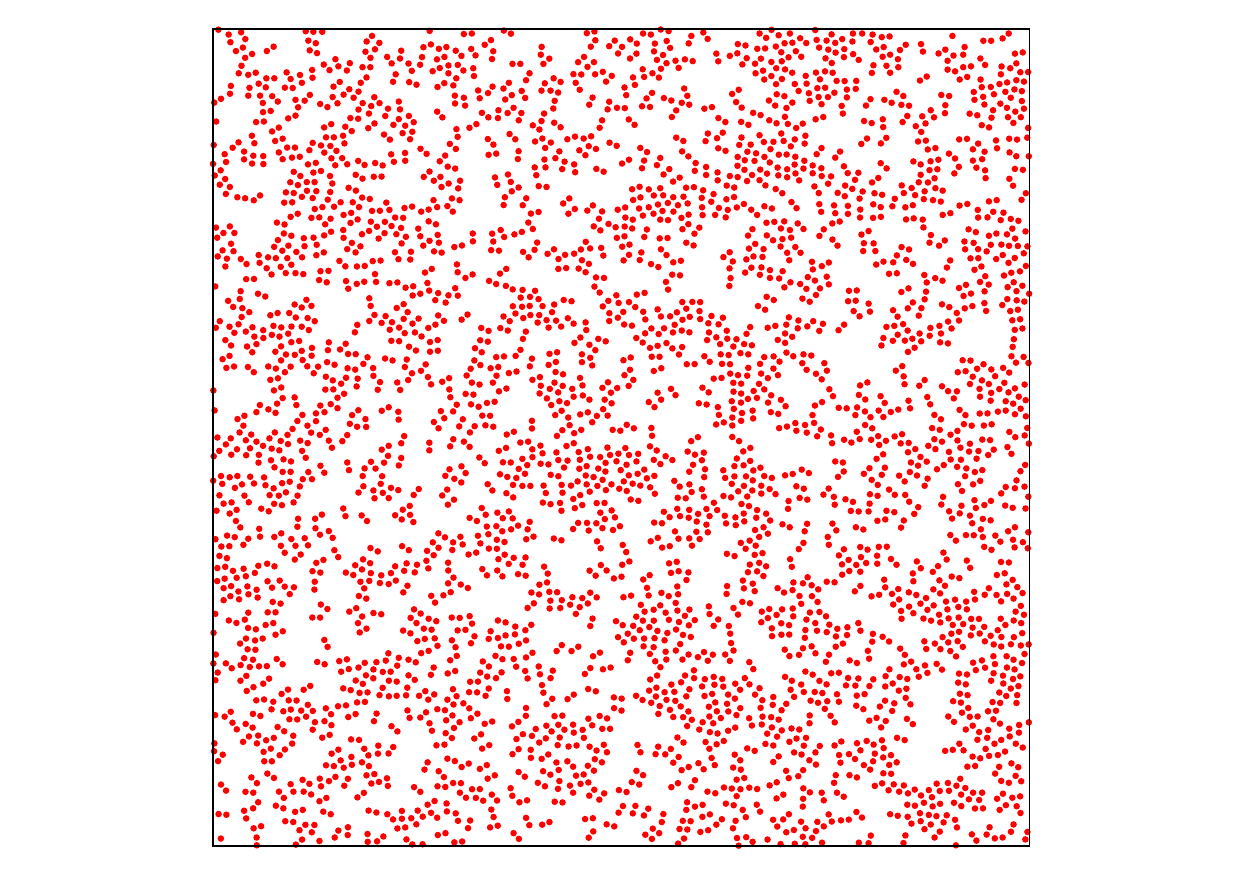}
        \hspace{-1cm}
     \includegraphics[scale=0.37]{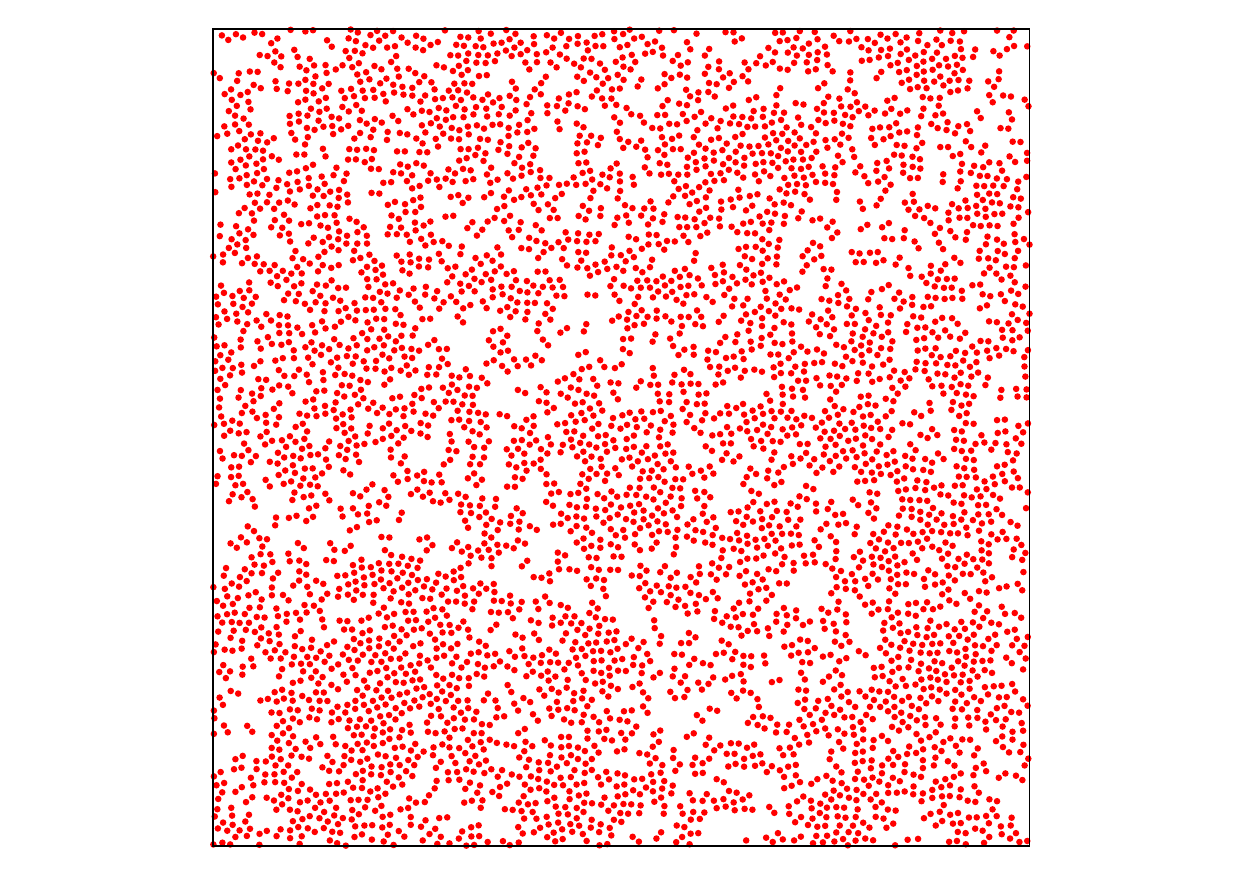}
          \\   %  (c)   \hspace{3.4cm}      (d)\\
 \end{center}
%    \\
%   \includegraphics[scale=0.34]{FIGS/vxdumbfi0_3}
\caption{(Color online.) 
Clustering effects in the active many-body system. In all snapshots the active force strength is $F_{\rm act}=0.5$ and the temperature is $T=0.05$,
while  the densities are $\phi=0.1, \ 0.2, \ 0.3, \ 0.4$ (in reading order). The side of the box is $l=100$.
}
    \label{fig:clusters_2}
\end{figure}

Figure~\ref{fig:clusters} shows six typical
snapshots of the system, all at the same density $\phi=0.4$
and temperature $T=0.05$
but for different active forces, $F_{\rm act}=0.01, \ 0.1, \ 0.3, \ 0.5, \ 0.7, \ 1$ (in reading order). 
The corresponding P\'eclet  numbers vary in the interval $[0.4,40]$
which is well inside the homogenous phase
in the phase diagram of Fig.~\ref{fig:phasediagram}.
On the 
%three
 four last snapshots, i.e. beyond $F_{\rm act} \simeq 0.3$ we start seeing clustering effects.
%These clusters, however, remain of finite constant size and are not stable, while in the aggregated phase they grow
%reaching the size of the system.
Their presence becomes less relevant moving away from
 the critical surface in parameter space.
%For the case with $F_{\rm act}=  1$ phase separation was found for $T \le T_c \approx 0.03$ corresponding to 
%Pe $\simeq 66.7$
%\cite{Suma13,Suma14}. For lower values of the P\'eclet number the system can  only be in a single phase. 
Configurations of the system for  different densities at fixed active force strength
 are  shown in Fig.~\ref{fig:clusters_2}.
 
\begin{figure}[b]
\begin{center}
  \begin{tabular}{cc}
    \includegraphics[scale=0.9]{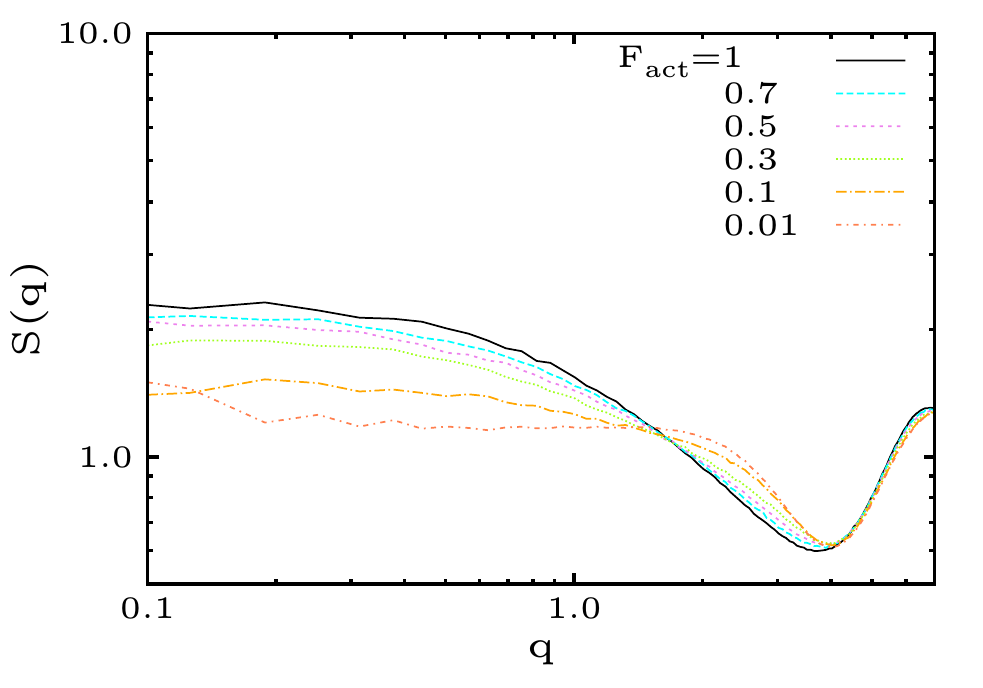}
    \\%(a)\\
    \includegraphics[scale=0.9]{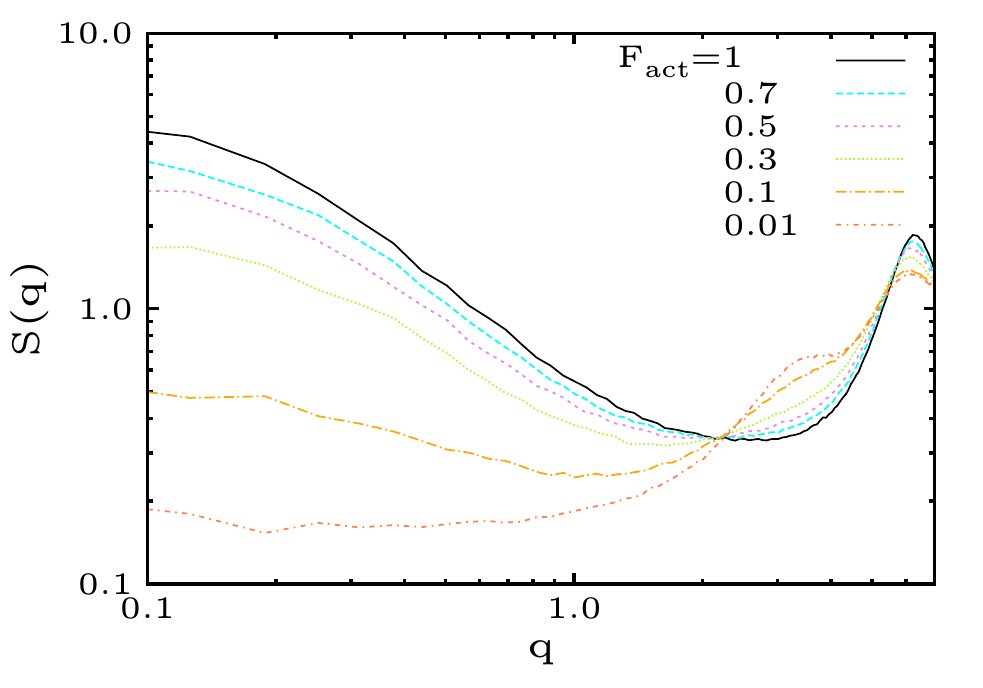}\\%(b)\\
 \end{tabular}
\caption{(Color online.) 
The structure factor of the active sample with $\phi=0.1$ (upper panel) and $\phi=0.4$ (lower panel)
for different active force strengths given in the key. See the text for a 
discussion.
}
    \label{fig:structure-factor-T0.05}
  \end{center}
\end{figure}

In Fig.~\ref{fig:structure-factor-T0.05} we show  the static structure factor
\begin{equation}
S(q) = \frac{1}{N} \sum_{i=1}^N \langle |e^{i {\mathbf q} \cdot {\mathbf r}_i} |^2 \rangle
\end{equation}
of a system with $\phi=0.1$ (upper panel) and $\phi=0.4$ (lower panel) for different active forces, 
$F_{\rm act} = 0.01, \ 0.1, \ 0.3, \ 0.5, \ 0.7, \ 1$
following the line  code  
given in the keys. The positions ${\mathbf r}_i$ here are taken to be the ones of the two beads in the diatomic molecule.
This function characterises the strength of density fluctuations at a length scale of the order of $2\pi/q$. 
Let us first discuss the data for $\phi=0.4$ (lower panel). 
For small active force strength, $F_{\rm act} = 0.01$, the system is very close to a simple fluid made of 
diatomic molecules. Consequently,  two peaks are visible in the curve. One represents the 
molecule elongation, i.e. the distance between the two beads in the dumbbell, and it is located at 
$q\simeq 6.21$ that corresponds to $\ell \simeq 1$. The other one signals the typical 
distance between beads belonging to different dumbbells and is located at $q\simeq 3.5$, that is to say $\ell \simeq 1.79$. 
For increasing values of the active force, clustering is favoured, and molecules in them tend to be closer to each other.
Therefore, the first peak (at longer distances in real space) progressively disappears while the second one (for distances
of the order of $\ell \simeq 1$) increases its weight. However, the 
most important new feature in the curves is the appearance and growth of the structure factor 
close to vanishing $q$. We note that the curves intercept at $q\simeq  2.26$ 
and that the curves move upwards with 
increasing $F_{\rm act}$ for $q$ smaller than this value. This increase quantifies the presence and growth of the clusters 
with increasing activity. 
The larger values of the structure factor at small $q$ observed for $F_{\rm act} \ge 0.3$ 
correspond to the clustering effects observed in this range of  $F_{\rm act}$
in Fig.~\ref{fig:clusters}.

Similar features in the structure factor were observed experimentally in~\cite{Theurkau12}
and numerically in~\cite{Fily12,Levis-Berthier,Lowen11}. In the dilute polymer melt active sample studied in~\cite{cugl-mossa3} no such important 
increase of the low-$q$ structure factor was observed and the conclusion was that 
active forces were making the sample more compact but uniformly, with no clustering effects.
The special $q$ values for the dumbbell system with packing fraction
$\phi=0.1$ are: the molecular elongation peak  is here located at $q \simeq 6.67$, the curves intercept at $q\simeq 1.48$ and
the second peak at $F_{\rm act}=0.01$ is not really visible. 

\begin{figure}[h]
\begin{center}
  \begin{tabular}{cc}
    \includegraphics[scale=0.9]{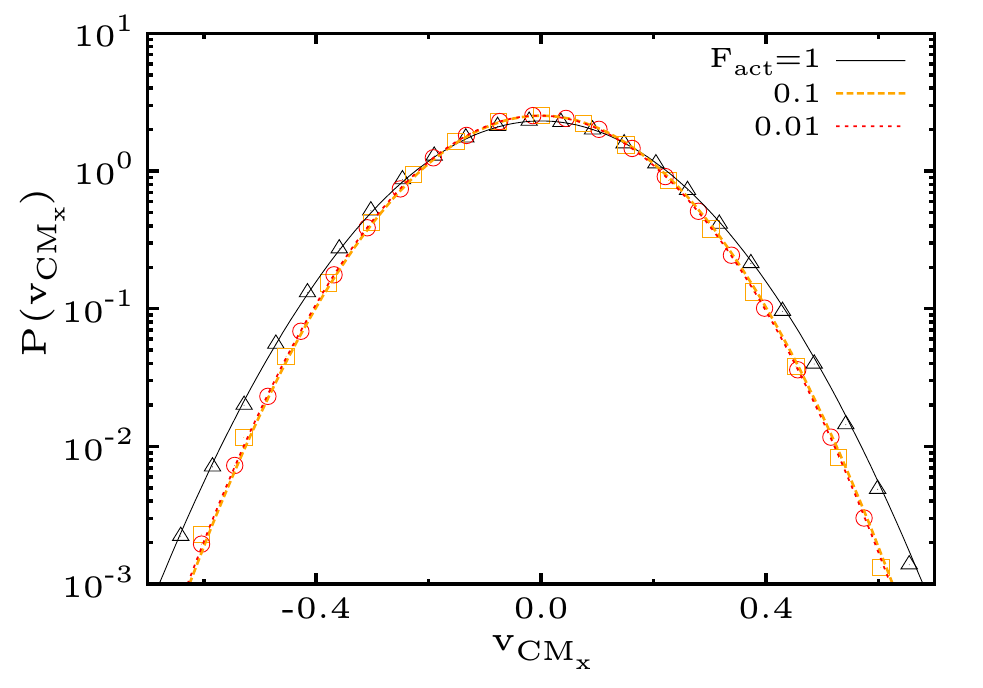}
%    \\
%    \includegraphics[scale=0.9]{FIGS/figure_lavoro_Antonio/fig12/vxdumbfi0_3}
  \end{tabular}
\caption{(Color online.) 
Pdf of the horizontal component of the center-of-mass dumbbell velocity, ${v_{CM}}_x$, 
in a system with $\phi=0.1$ at $T=0.05$, for three values
of the active force, $F_{\rm act}=0.01, \ 0.1, \ 1$.  Data are shown in linear-log scale. 
The values of the kinetic temperature, $T_{\rm kin} = 0.0500, \ 0.0501, \  0.0596$, respectively, 
used in the Gaussian fit shown with continuous and dashed lines 
are very close to the result for a single molecule,  see  Eq.~(\ref{kinetic_temperature_formula}).
At $\phi=0.3$, for the same set of active forces, we have checked that the pdf is also well represented
by a Gaussian distribution with $T_{\rm kin}$ given by Eq.~(\ref{kinetic_temperature_formula}). 
%\approx 0.0489, \ 0.0489, \  0.0523$ respectively.
}
    \label{pdfdumb}
  \end{center}
\end{figure}

Although we have not shown it analytically, the probability distribution functions (pdf) of the velocity components
is well represented by a Gaussian pdf with a kinetic temperature given by 
Eq.~(\ref{kinetic_temperature_formula}) for small values of $\phi$, as long as phase separation does not 
occur. These pdfs are shown with continuous and dashed lines in Fig.~\ref{pdfdumb}. 
Increasing the surface concentration, jamming effects 
slightly reduce the overall $T_{\rm kin}$.

In essentially out of equilibrium systems with non-potential forces that drive the dynamics, such as driven granular matter~\cite{Jaeger96,Aronson06,Pouliquen08}, 
vortex lattices~\cite{kolton} and the active matter we here study, the kinetic temperature should be higher than the ambient 
temperature. In our case we measured $T_{\rm kin}\simeq 0.05$ for $F_{\rm act} = 0.01$ and
%$T_{\rm kin}=0.05$ for 
$F_{\rm act}=0.1$,
% We conclude that the numerical accuracy cannot be such to trust the last digit and 
so that
$T_{\rm kin} \simeq T$ in these two cases.
 For the strongest $F_{\rm act}$ used, $F_{\rm act}=1$, 
we measured $T_{\rm kin} = 0.0596>T$. We can compare this value to the one expected for the single active dumbbell under the 
same conditions, according to Eq.~(\ref{kinetic_temperature_formula}): $T_{\rm kin}^{\rm single} \simeq 0.06$ 
which is very close to the value 
measured for the ensemble.  

\begin{figure}[h]
\begin{center}
  \begin{tabular}{cc}
%   \includegraphics[scale=0.9]{FIGS/figure_lavoro_Antonio/fig13/diffusion-constant-active-dumbbells-different-Fa_fi_constant.eps}
%   \\
    \includegraphics[scale=0.9]{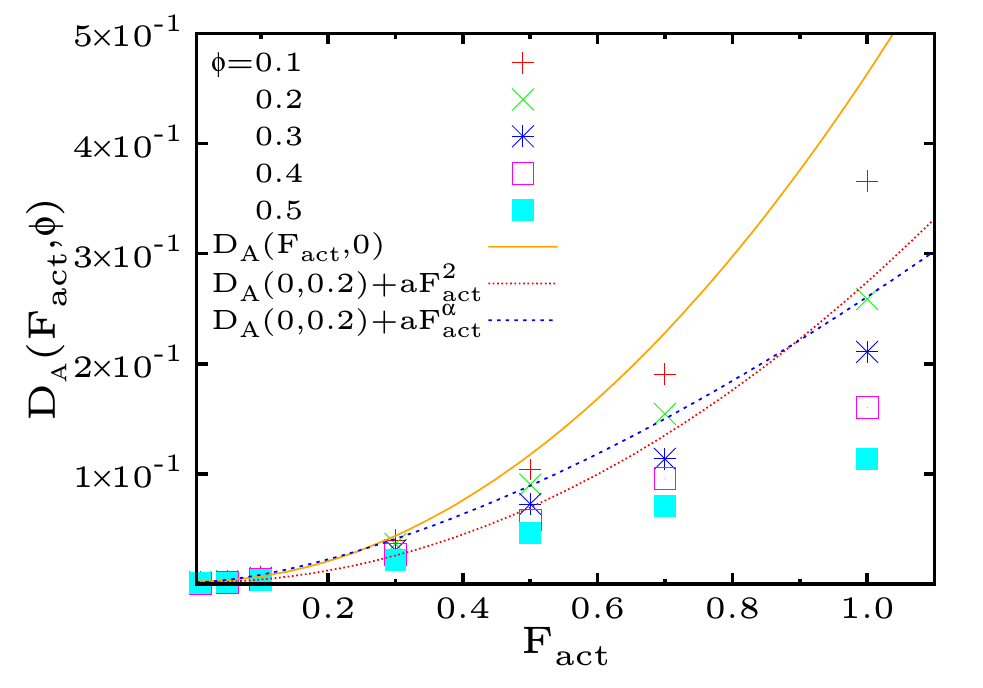}
 \end{tabular}
\caption{(Color online.) 
The diffusion constant in the active system at $T=0.05$, 
$D_A(F_{\rm act}, \phi)$, as a function of $F_{\rm act}$ for five values of the system's surface fraction  given in the key.
Normal scale is used. 
The continuous (orange) line is the analytic expression for $\phi=0$, see Eq.~(\ref{active_diffusion_constant}). 
The dashed (red) line is a fit of the data
to the form $D_A(0,0.2)+a F_{\rm act}^2$ with $D_A(0,0.2)=0.00153$ obtained from the passive data at this 
temperature and $a=0.273$ (the $F^2_{\rm act}$ 
dependence found for the single molecule)
for $\phi=0.2$ (green crosses $\times$). 
The dashed (blue) line is a fit to the form $D_A(0,0.2)+a F_{\rm act}^\alpha$ with $D_A(0,0.2)=0.00153$, $a=0.259$ 
and $\alpha\simeq 1.56$
(the fit used in~\cite{cugl-mossa1,cugl-mossa3}, though $\alpha\simeq 2.3$ was found in that case for the low density used).
The corrections to this fit will be studied in the next figures and discussed in the text. 
}
    \label{diff_active_dumb_T0_05-b}
  \end{center}
\end{figure}

\subsection{Mean-square displacement and asymptotic diffusion}

\subsubsection{Very low temperature}

In order to appreciate the effects of the dumbbell interaction,
in Fig.~\ref{fig:dumbattivevarifi_1} we showed with  thicker lines 
the mean-square displacement in the active dumbbell 
%interacting
 system with $\phi=0.1$ and the 
same four values of the active force $F_{\rm act}=0.001, \ 0.01, \ 0.1, \ 1$ used in the single dumbbell case. 
%The colour code is such that
Data  for single and collective systems under the same active force are shown with the same
color and line style. For the two smaller applied forces,  $F_{\rm act}=0.001, \ 0.01$, we do not see 
any difference in $\Delta^2$ between the 
single and interacting case in this scale. For the two larger forces, $F_{\rm act}=0.1, \ 1$, 
the interaction between the molecules slows down the dynamics in the sense that the 
mean-square displacement of the interacting system, at the same time-lag, is smaller than the 
one for the single molecule. For the forces $F_{\rm act}=0.001,\ 0.01, \ 0.1$ we still see
the angular time scale $t_a$, a feature that does not exist in models of point-like active particles 
as the ones studied in~\cite{cugl-mossa1,cugl-mossa3,Levis-Berthier}.
Note that for $F_{\rm act}=1$ and such a low temperature, $T=0.001$, the
system is in the phase-separated phase, the velocity-component pdf is no longer Gaussian (not shown) and 
the mean-square displacement is much slower than in the single molecule case.

A first presentation of the diffusion constant, $D_A(F_{\rm act},\phi)$ obtained by studying the system
at the different surface fractions at  a very low temperature, $T=0.001$, and for the active forces,
 $F_{\rm act}=0.1, 0.001$,
was given in Fig.~\ref{fig:diffusive-constant-passive-dumbbells}.
In this plot $D_A(F_{\rm act},\phi)$ was normalized by the 
center-of-mass diffusion constant of a single active molecule under the same conditions, $D_A(F_{\rm act},0)$. 
At $F_{\rm act}=0.1$, the  
data fall-off from  1 at $\phi=0$ very quickly as soon as $\phi>0$. At this very low temperature and 
relatively high active force small clusters appear and the diffusion properties are altered by them.
On the other hand, at $F_{\rm act}=0.001$ the data are close to those for the passive system but they lie above them. 
They are not well represented by the TO expression and, moreover, they give a first indication of non-monotonic
dependence of the normalised data with the active force, that we will discuss in more detail below when using a 
higher bath temperature.

%\subsection{Dynamical properties and effective temperature}

\subsubsection{Working temperature}

We  will show  in the following results for a higher temperature,   $T=0.05$, such that, for all the active forces considered,
the system is in the homogeneous phase, even though, as showed before, non--trivial  fluctuation effects are clearly observable
at sufficiently high values of  $F_{\rm act}$. We  also studied the system at other higher temperatures obtaining results  similar to those  found at $T=0.05$.  
(The values of the effective temperature found at $T=0.1$ will be  reported in Table 1.)

The diffusion constant in the last diffusive regime of the active system is shown, as a function of active force, 
in Fig.~\ref{diff_active_dumb_T0_05-b} for five densities $\phi=0.1, \ 0.2, \ 0.3, \ 0.4, \ 0.5$.
$D_A(F_{\rm act},\phi)$ was  extracted from the analysis of the mean-square displacement in the 
late $t\gg t_a$ regime. The figure demonstrates that $D_A(F_{\rm act},\phi)$ decreases with increasing $\phi$ and increases 
with  increasing $F_{\rm act}$. The analytic expression for the single particle limit, $\phi=0$, given in 
Eq.~(\ref{active_diffusion_constant}) is added with a continuous (orange)  line. Clearly, the finite density induces a
reduction of the diffusion constant for all $F_{\rm act}$. A quadratic fit of the data for $\phi=0.2$, which conserves the 
$F_{\rm act}^2$ dependence of the single particle limit, is shown with a 
dashed (red) line. More precisely, 
the form  used is $D_A(0,0.2)+a F_{\rm act}^2$ with $D_A(0,0.2)=0.00153$ obtained from the passive data at this 
temperature and $a=0.273$
for $\phi=0.2$ (green crosses $\times$). Had we left the $F_{\rm act}=0$ intercept as a free parameter in this fit we would have obtained $0.0098$ which is relatively close  to the actual passive limit, as it should.  
This fit is acceptable at small $F_{\rm act}$ but a deviation at large values of the active force is clear. 
Another fit, with an additional fit parameter as the power in the $F_{\rm act}$ dependence, is also 
shown with a dashed (blue) line. This fit was used in~\cite{cugl-mossa1,cugl-mossa2} to describe the 
diffusion constant of a rather dilute ensemble of interacting active point-like particles. 
We find here that a power law with $\alpha\simeq 1.56$ fits the dumbbell data at $\phi=0.2$ rather correctly for all 
$F_{\rm act}$ shown. The analysis of the $\phi$-dependence will be done next, where a more convenient way
of describing these data will be proposed.

\begin{figure}
\begin{center}
  \begin{tabular}{cc}
%\includegraphics[scale=0.35]{FIGS/mean-square-displacement-attive-dumbbells-diffusion-fi0_1-T0_05}
%    \\
    \includegraphics[scale=0.9]{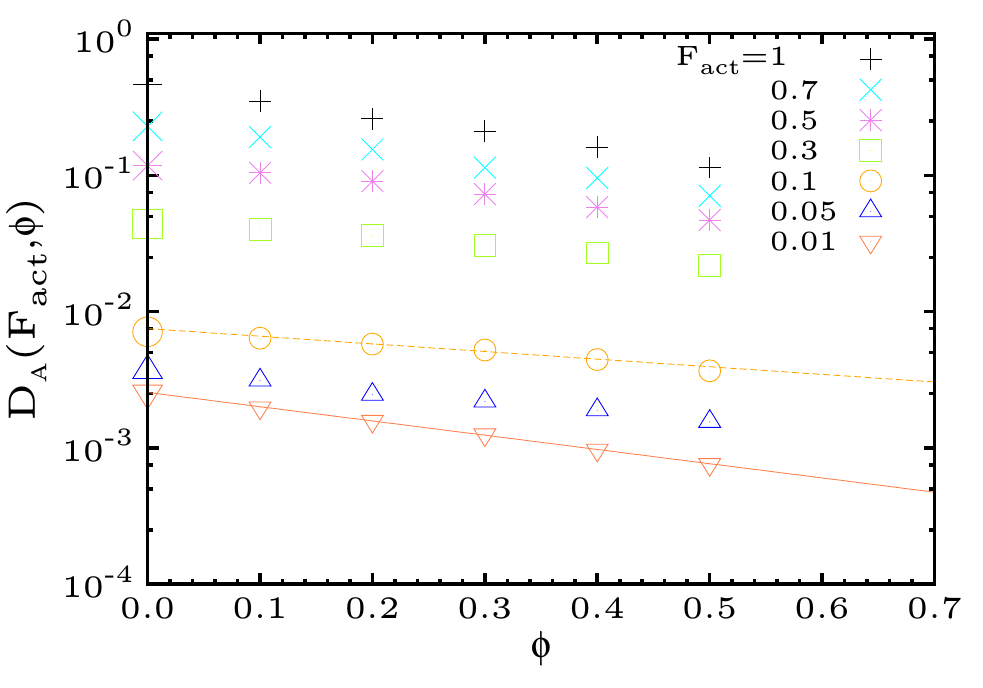}
    \\%(a)\\
%    \includegraphics[scale=0.9]{FIGS/figure_lavoro_Antonio/fig13/diffusion-constant-active-dumbbells-different-fi-Fa-scalanormale.eps}
%    \\
     \includegraphics[scale=0.9]{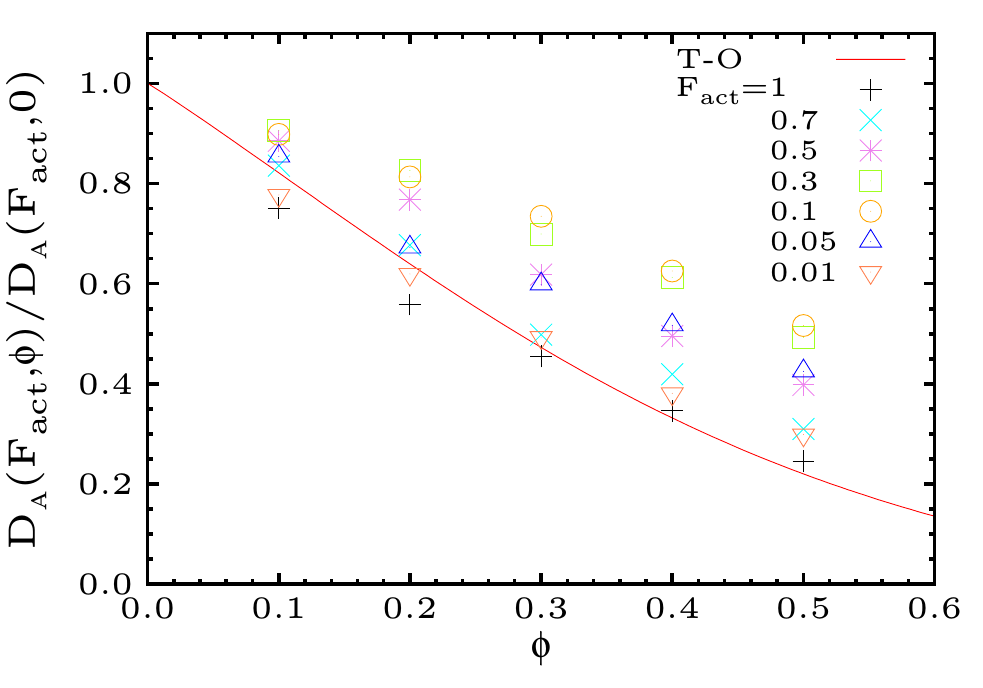}\\%(b)\\
  \end{tabular}
\caption{(Color online.) 
The diffusion constant, as extracted from the 
mean-square displacement of the dumbbells center-of-mass, as a function of 
$\phi$, in a system of active molecules at $T=0.05$. 
Upper panel:  A linear-ln scale is used due to the large
 variation in the absolute values of $D_A(F_{\rm act}, \phi)$. The values of $F_{\rm act}$ are given in the key.
  The values at $\phi=0$, shown with  larger symbols, correspond to the theoretical value of $D_A$ 
 for a single dumbbell given by Eq.~(\ref{active_diffusion_constant}). 
 The data for relatively small $\phi$ are rather well fitted by a straight line, suggesting 
 $\ln D_A(F_{\rm act}, \phi)= \ln D_A(F_{\rm act}, 0) -b(F_{\rm act})\phi$ with $D_A(F_{\rm act}, 0)$ the single active molecule diffusion constant.
In the lower panel the data are normalised by the diffusion constant of the single molecule and they are presented in linear scale.
Non-monotonic behaviour with respect to $F_{\rm act}$ is observed suggesting that the fitting function $b(F_{\rm act})$ should be non-monotonic.
}
    \label{diff_active_dumb_T0_05}
  \end{center}
\end{figure}

\begin{figure}
\begin{center}
  \begin{tabular}{cc}
     \includegraphics[scale=0.9]{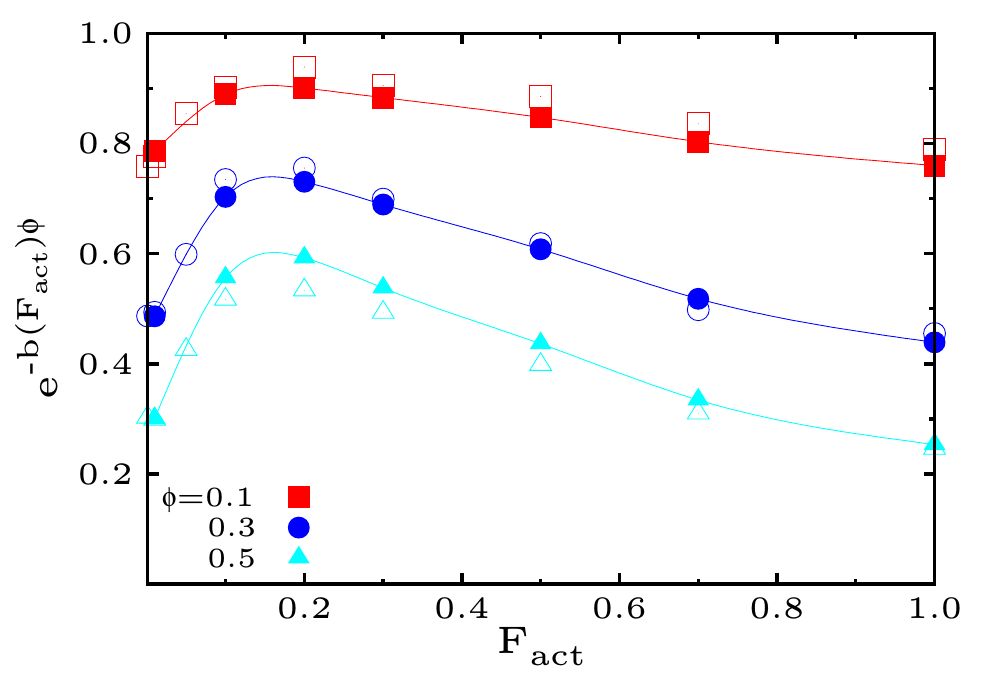}
     \\%(a)\\
%\includegraphics[scale=0.35]{FIGS/mean-square-displacement-attive-dumbbells-diffusion-fi0_1-T0_05}
%     \includegraphics[scale=0.9]{FIGS/figure_lavoro_Antonio/fig13/fig3/b.eps}
%\\
    \includegraphics[scale=0.9]{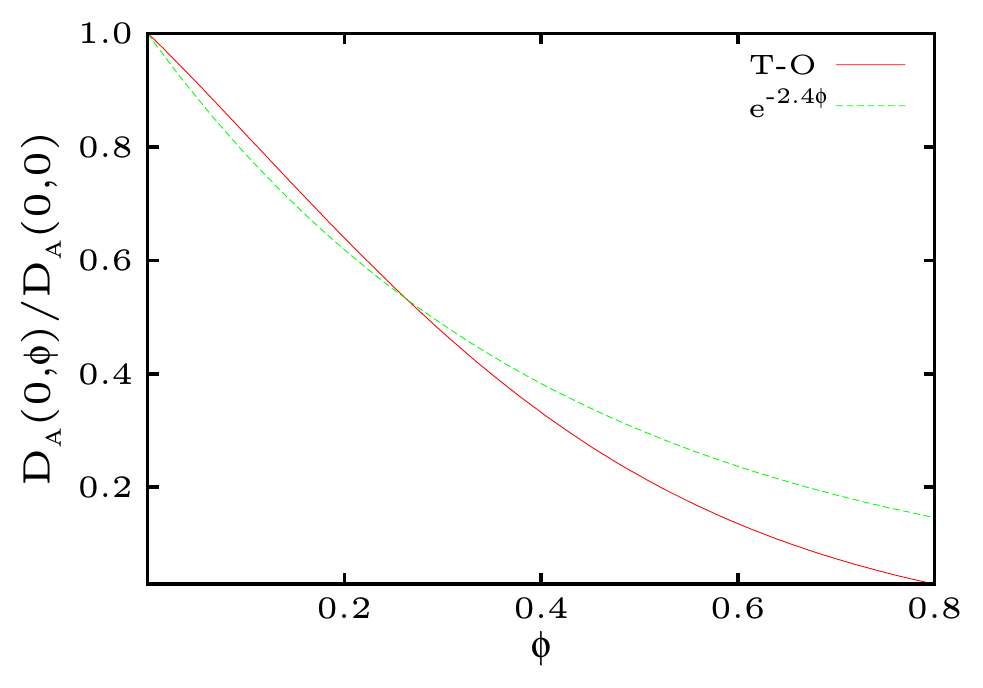}\\%(b)\\
     \end{tabular}
\caption{(Color online.) 
Upper panel:  the exponential $e^{-b(F_{\rm act}) \phi}$ 
 for three densities given in the key. Original data are represented by open symbols
 while the filled symbols  represent the values obtained from 
 the fit in Eq.~(\ref{diff_different_fi_approx})
of $\frac{D_A(F_{\rm act},\phi)}{D_A(F_{\rm act},0)}$
performed in Fig.~\ref{diff_active_dumb_T0_05}.
The lines are guides-to-the-eye obtained with a spline of the filled points.  
Lower panel: the correct TO description of data, $\frac{D_A(0,\phi)}{D_A(0,0)}=[1+H(\phi)]^{-1}$
for $F_{\rm act}=0$ (with red solid line), confronted to the exponential approximation in Eq.~(\ref{diff_different_fi_approx}), 
that in this case yields $b(F_{\rm act}=0)=2.4$ (with green dashed  line). 
The two curves are very close to each other up to $\phi\lesssim 0.3$.
}
    \label{confronto_exp_TH}
  \end{center}
\end{figure}

In Fig.~\ref{diff_active_dumb_T0_05} we show the same diffusion constant, $D_A(F_{\rm act},\phi)$,
as a function of the surface fraction $\phi$.
The values of the  active force $F_{\rm act}$ are given in the key. 
We have already stressed, when showing the very low temperature data, that as
soon as $F_{\rm act}\neq 0$ the rather complex TO packing fraction dependence breaks down. We 
therefore seek for a different (and simpler) $\phi$-dependence of the diffusion constant of  the active interacting sample.
In the linear-ln presentation the data (up to $\phi\simeq 0.4$) are rather well fitted by a straight line, suggesting 
 \begin{equation}
 D_A(F_{\rm act}, \phi) \simeq D_A(F_{\rm act},0)  \ e^{-b(F_{\rm act}) \phi}
 \label{diff_different_fi_approx}
 \end{equation}
 with $D_A(F_{\rm act}, 0)$ the single active molecule diffusion constant. (This is confirmed by a double linear presentation of data.)
 The exponential dependence on $\phi$ is much simpler than the 
 TO expression and it cannot be taken as a formal proposal for the behavior of $D_A$ as we do not have an analytic 
 justification for it. We simply stress
 here that it provides an acceptable description of data at this temperature.
 
In the second panel in Fig.~\ref{diff_active_dumb_T0_05} the finite $\phi$ diffusion constant is normalised by the single molecule one, 
in the form used in Fig.~\ref{fig:diffusive-constant-passive-dumbbells}. The plot shows an unexpected
 non-monotonic behavior as a function of $F_{\rm act}$ suggesting that the effective $b(F_{\rm act})$ should be non-monotonic.
It is interesting to observe that the diffusion constant ratio increases at small values of 
 $F_{\rm act}$ starting to decrease when clustering effects become relevant for $F_{\rm act} \ge 0.3$.

In the upper panel in  Fig.~\ref{confronto_exp_TH} we display the exponential factor $e^{-b(F_{\rm act}) \phi}$. We use 
open symbols
for the result of the ratio $D_A(F_{\rm act}, \phi)/D_A(F_{\rm act}, 0)$ between numerical data, and filled symbols  for the 
 the fit of $D_A(F_{\rm act}, \phi)/D_A(F_{\rm act}, 0)$,
as a function of $F_{\rm act}$.  Three values of $\phi$ were used and are given in the key.
Extracting $b(F_{\rm act})$ from here one confirms the non-monotonic dependence with $F_{\rm act}$ with  
$b$ varying  in the interval $ [1, \ 2.7]$, circa. One observes that the fit proposed in Eq.~(\ref{diff_different_fi_approx})
works reasonably well for all  the values considered for  $F_{\rm act}$.
The approximate exponential fit of data is compared to the TO prediction 
for $F_{\rm act}=0$ in the lower panel. The two are very close for $\phi \lesssim 0.3$ while they deviate 
considerably for higher density. 

To the best of our knowledge this non-monotonic behaviour has not been 
observed in the literature yet. 

\subsection{The linear response function}

We now turn to the study of the linear response function, integrated over a period of time, as defined in 
Eq.~(\ref{eq:def-chi}). 

\begin{figure}[h]
\begin{center}
%\includegraphics[scale=0.9]{FIGS/figure_lavoro_Antonio/fig14/chi-dumbbells-phi_1-T_05.eps}
% \\
    \includegraphics[scale=0.9]{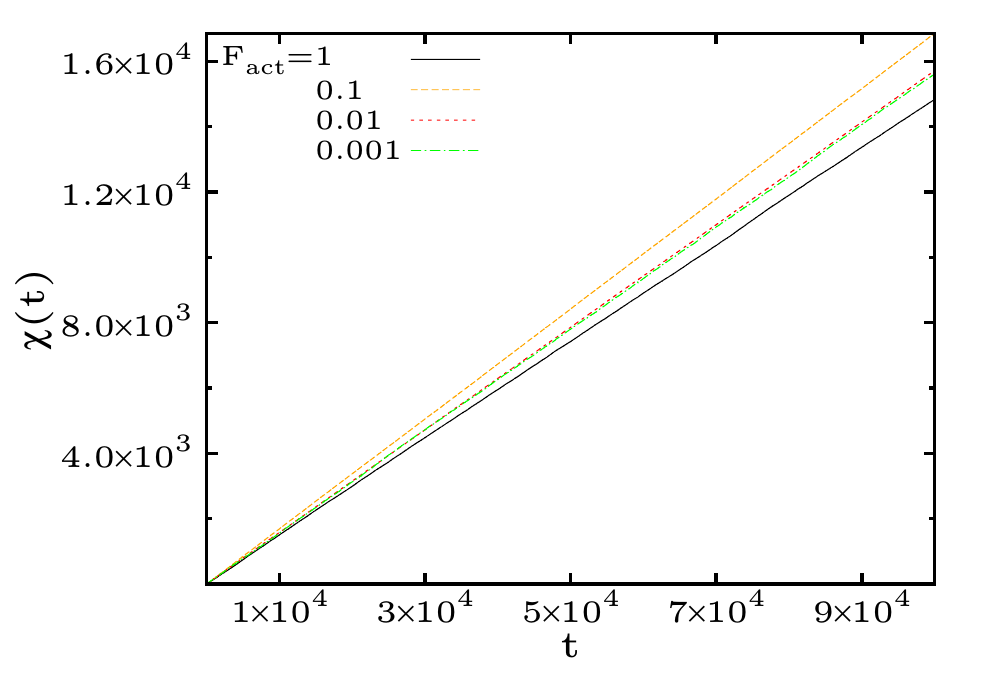}
    \\%(a)\\
    \includegraphics[scale=0.9]{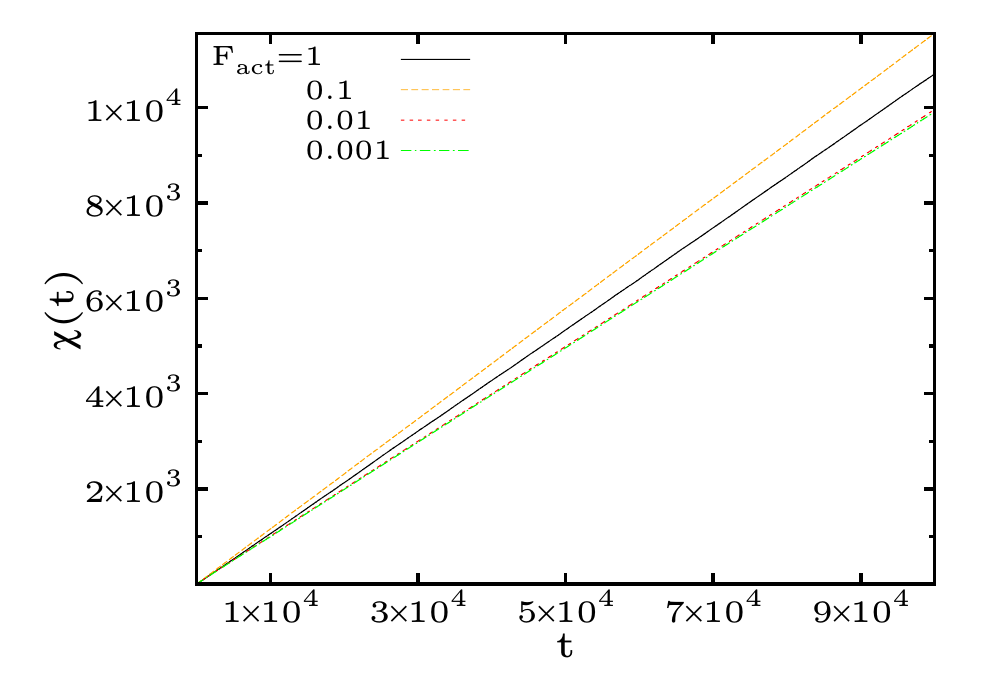}\\%(b)\\
%\centerline{\epsfig{file=diff-tot+storie7.eps,width=0.7\textwidth,angle=-90,clip=}}
\caption{(Color online.) 
The induced displacement or linear susceptibility 
for a system of active dumbbells at $T=0.05$ and surface fractions $\phi=0.1$ (above) and $\phi=0.3$ (below).
The different values of the
active force used are given in the key. The magnitude of the applied field is 
$f=0.01$.
 One sees a non-monotonic dependence of $\mu(F_{\rm act},\phi)$ on $F_{\rm act}$, similarly to 
 what we observed in $D_A(F_{\rm act}, \phi)/D_A(F_{\rm act},0)$.  
  }
\label{fig:chi-active}
\end{center}
\end{figure} 

We started with an analysis of the 
amplitude of the applied perturbation, so as to determine the optimal value to be used for each set of parameters.
Indeed, one has to use a small enough applied force to keep the response within the linear regime, but large enough 
to reduce the fluctuations. We found that $f=0.01$ yields the best results. 

\begin{figure}[h]
\begin{center}
\includegraphics[scale=0.9]{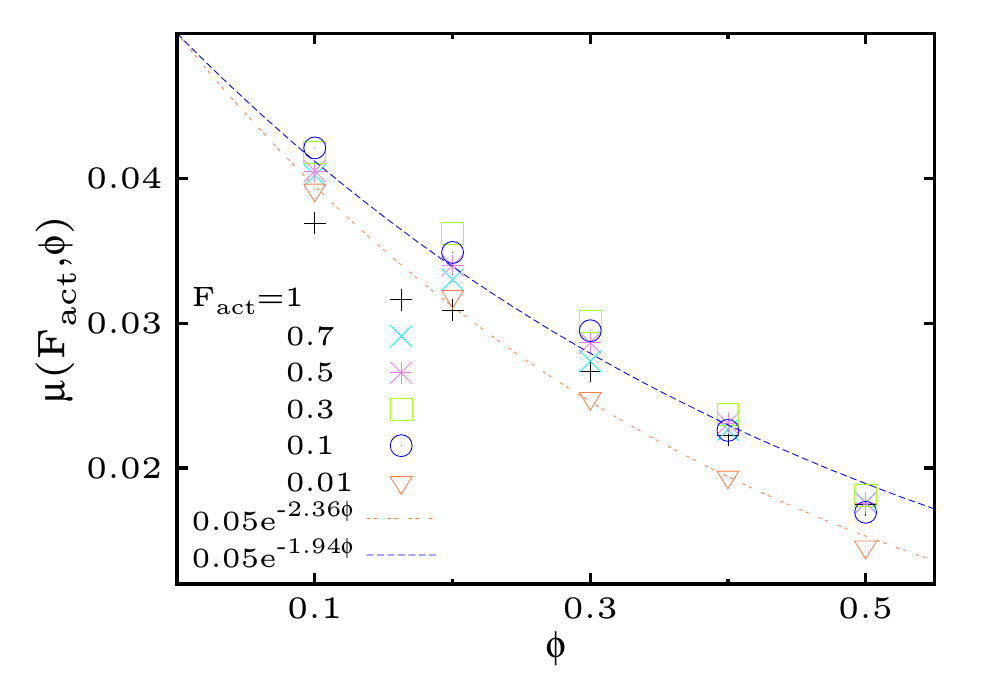}
\\%(a)\\
%\includegraphics[scale=0.9]{FIGS/figure_lavoro_Antonio/fig14/chi-dumbbells-prop-constant_rapporto.eps}
%  \\
% \includegraphics[scale=0.9]{FIGS/figure_lavoro_Antonio/fig14/plot_chi_constant_Fa/chi-dumbbells-prop-constant.eps}
%   \\
 \includegraphics[scale=0.9]{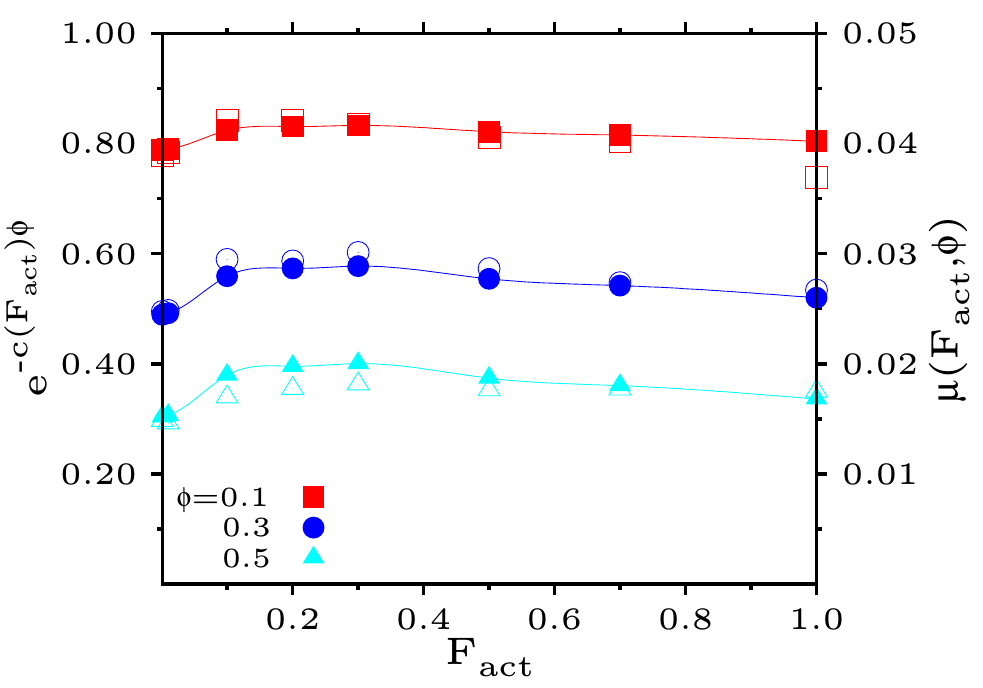}\\%(b)\\
%\centerline{\epsfig{file=diff-tot+storie7.eps,width=0.7\textwidth,angle=-90,clip=}}
\caption{(Color online.) 
Upper panel:
The asymptotic slope of the integrated linear response
$\mu(F_{\rm act},\phi)$
as a function of $\phi$ for various active forces given in the key. The two curves are 
exponential fits to the data for $F_{\rm act} = 0.01, \ 0.1$; the pre factor is $1/(2\gamma)=0.05$.
Lower panel: the ratio $\mu(F_{\rm act}, \phi)/\mu(F_{\rm act}, 0)$
 is plotted as a function 
of $F_{\rm act}$ for the density values given in the key. The scale on the right side gives the values 
of $\mu(F_{\rm act}, \phi)$.
 With open symbols, the ratios between numerical data; with 
filled symbols the result of the exponential fit of data. The curves are non-monotonic and the relative motility is enhanced for active 
forces of the order of $0.1 \lesssim F_{\rm act} \lesssim  0.4$. 
}
\label{fig:chi-active_other}
\end{center}
\end{figure}

Figure~\ref{fig:chi-active} shows the time integrated linear response function $\chi(t)$ as a function of time $t$
for various active many-body systems 
at $T=0.05$ and $F_{\rm act}=0.001, \ 0.01, \ 0.1, \ 1$. 
In the two panels ($\phi=0.1$ and $\phi=0.3$) double linear scales are used 
and the dependence of the linear response integrated over time
on $F_{\rm act}$  is now made visible. 
The induced displacement at a given time $t$  decreases   with the dumbbells' 
concentration. While the $F_{\rm act}=0.001$ and $F_{\rm act} =0.01$ curves lie on top of each other 
(within numerical accuracy), we notice the non-monotonic dependence on $F_{\rm act}$ for higher 
activities. Indeed, in both panels the 
curve for   $F_{\rm act}=0.1$ lies above the ones for $F_{\rm act} = 0.001, \ 0.01$ and $F_{\rm act}=1$.
Moreover, the data for $F_{\rm act}=0.01$ and $F_{\rm act} = 1$ appear in different order
for $\phi=0.1$ and $\phi=0.3$. We have not included data for intermediate active forces $0.1 < F_{\rm act} < 1$
in this plot to ease the visualization but we have analysed them to extract the asymptotic slope (see the data 
presented in Fig.~\ref{fig:chi-active_other} below).

The non-monotonic dependence of $\chi(F_{\rm act}, \phi)$ on $F_{\rm act}$
for sufficiently large $\phi$  is  reminiscent of -- though not the same as -- 
a negative resistivity, that is to say a decreasing dependence of a current on the 
applied field that drives it,  observed, for instance, in sufficiently dense kinetically constrained 
systems~\cite{Sellitto08}.

We extract the slope of these time-dependent curves  for times such that $t\gg t_a$ 
and we call it $d \mu(F_{\rm act}, \phi)$. We already know that at fixed $F_{\rm act}$, it is 
a decreasing function of the dumbbell concentration $\phi$, and that at fixed density it is a
non-monotonic function of $F_{\rm act}$.

 The study of the dependence of $\mu$ on these two 
parameters is performed in Fig.~\ref{fig:chi-active_other}. 
In the upper panel we plot $\mu(F_{\rm act}, \phi)$ as a function of $\phi$ for several 
active forces given in the key. We confirm the 
 monotonic decay with increasing density. 
The two dashed curves are exponential fits as function of $\phi$ for two choices of the 
active force, $F_{\rm act}=0.01, \ 0.1$.
 
 However, the dependence on $F_{\rm act}$ is less straightforward. The curves for different $F_{\rm act}$ 
 cross. For instance, there is an inverting value of $\phi$,
say $\phi^*$,  so that  $\mu(F_{\rm act}=1, \phi<\phi^*)$ is smaller that 
$\mu(F_{\rm act}=0.1, \phi<\phi^*)$ while one observes 
  $\mu(F_{\rm act}=1, \phi>\phi^*) >
\mu(F_{\rm act}=0.1, \phi>\phi^*)$.  
At fixed density, the  relative motility is enhanced for active 
forces of the order of $0.1\lesssim F_{\rm act} \lesssim   0.4$ but the trend is reversed for
higher active forces.  The maximal effect is seen
for $\phi\simeq 0.2$.

In the second panel we show the $F_{\rm act}$-dependence of the ratio $\mu(F_{\rm act}, \phi)/\mu(F_{\rm act}, 0)$
 for three packing fractions given in the key.
The factor $\mu(F_{\rm act}, 0)=1/(2\gamma)$ is just a constant independent of $F_{\rm act}$ and $\phi$. 
We use open  data points to represent the ratio between the numerical data, and filled symbols with 
joining lines to represent the outcome of the fit to an  exponential:
\begin{equation}
\mu(F_{\rm act}, \phi) = \mu(F_{\rm act}, 0) \ e^{-c(F_{\rm act}) \ \phi}
\; . 
\end{equation}
 Note that if 
$c(F_{\rm act})$ were given by $b(F_{\rm act})$ 
as for the diffusion constant, we would have a $\phi$-independent
$T_{\rm eff}$, and the same as for $\phi=0$. 
We obtain a non-monotonic dependence on $F_{\rm act}$ for $\phi \lesssim  0.4$ while the 
data for the highest density, $\phi=0.5$, may approach a plateau at the value reached at $F_{\rm act}=0.1$.

\subsection{Fluctuation-dissipation relation}

\begin{figure}[h]
\begin{center}
  \begin{tabular}{cc}
    \includegraphics[scale=0.9]{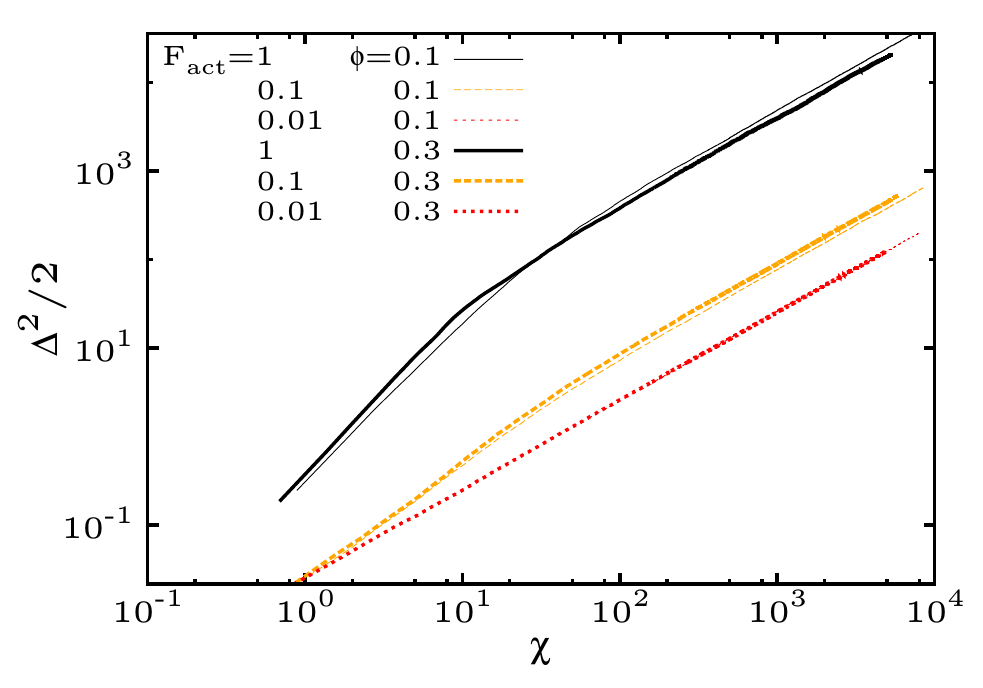}
    \end{tabular}
    \caption{(Color online.) 
    Parametric plot $\Delta^2(\chi)$ for three values of the activity, $F_{\rm act}=0.01, \ 0.1, \ 1$, 
two values of the surface concentration $\phi=0.1, \ 0.3$, and $T=0.05$. Double logarithmic 
representation. The effective temperature can be read from the off-set in the $y$ direction of the projection of the straight line in the 
long-time regime after the shoulder, $\ln T_{\rm eff} = \ln \Delta^2 - \ln 2\chi$. }
    \label{T_eff_parametricplot}
      \end{center}
    \end{figure}

In Fig.~\ref{T_eff_parametricplot} we display the parametric plot $\Delta^2(\chi)$ for the three choices of activity, 
$F_{\rm act}=0.01, \ 0.1, \ 1$ and two surface fractions $\phi=0.1, \ 0.3$ all at 
$T=0.05$. We see a very weak dependence on $\phi$ and a strong dependence on
$F_{\rm act}$. All data points at $F_{\rm act}=0$ (not shown) fall on top of each other (independently of 
$\phi$) as the passive system is 
in equilibrium with the bath. The data for $F_{\rm act}=0.01$ are very close to these and the data fit 
gives values of the effective temperature (the exponential of the vertical axis off-set of the slope of the curves) 
that are near the ambient temperature as well (see the numerical estimates  in Table~\ref{tab:template}).

For larger values of the activity, the parametric construction displays the familiar shoulder separating short from 
long time-delay behaviour~\cite{cugl-kur-pel,cugl:review}. The short time-delay fluctuations are much affected by 
the microscopic dynamics controlled, in this case, by the ambient temperature (the same for all sets of curves) and the 
activity (very different for the three sets of curves shown with different colour and line style). 

\begin{figure}[h]
\begin{center}
  \begin{tabular}{cc}
    \includegraphics[scale=0.9]{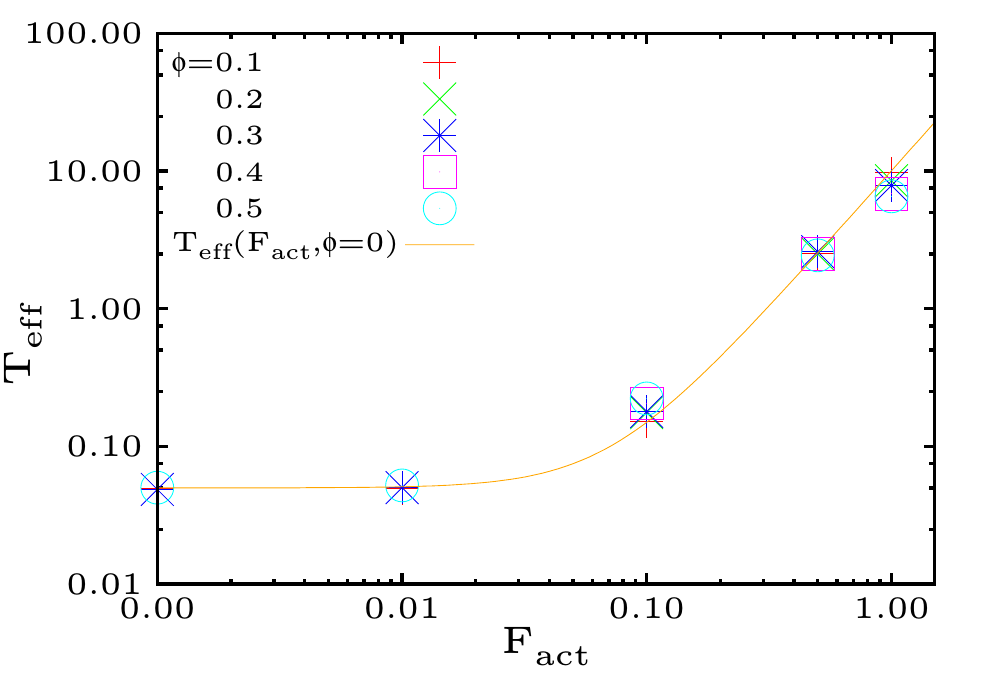}
  \end{tabular}
\caption{(Color online.) 
Fitted values of $T_{\rm eff}$ from the parametric plot in Fig.~\ref{T_eff_parametricplot} 
for five values of $\phi$ given in the key,  all at the  temperature $T=0.05$. In this representation, all data seem to be in good agreement with
the $T_{\rm eff}$ formula for a single dumbbell, Eq.~(\ref{T_eff_equation}), that is represented with the solid line.
However, the actual values of $T_{\rm eff}$ are given in Table~\ref{tab:template} where one sees that there is, though, 
a weak difference in the data for different $\phi$:
for $F_{\rm act}=0.1$, $T_{\rm eff}$ tends to increase with 
increasing $\phi$, while for $F_{\rm act}=1$, $T_{\rm eff}$ tends to decrease with increasing $\phi$. 
}
  \label{fitted-values}
    \end{center}
\end{figure}

After the shoulder, at long time-delays, 
 the structural dynamics where interactions between dumbbells are important sets in. The effective temperature will be 
extracted from the slope of the parametric curves (in linear scale) in this time-delay regime, or from the vertical axis 
off-set of the slope in the late-time regime in double logarithmic scale, as explained above.
That is to say, 
the asymptotic relation between linear integrated response and mean-square displacement 
allows one to extract the effective temperature as
\begin{equation}
k_BT_{\rm eff}(F_{\rm act}, \phi)  = 
\frac{D_A(F_{\rm act},\phi)}
       {\mu(F_{\rm act},\phi)} 
\; . 
\end{equation}
The effective temperature values are larger than the environmental temperature for $F_{\rm act} >0$. 
The numerical dependence of $T_{\rm eff}$ on $F_{\rm act}$ is reported in Fig.~\ref{fitted-values}, where 
$T_{\rm eff}$ against $F_{\rm act}$ is shown. Note that $T_{\rm eff}$ is consistently larger than 
$T$ and increases with $F_{\rm act}$. The solid line in the plot represents the theoretical 
result for a single active dumbbell, $T_{\rm eff} \simeq T+ c F_{\rm act}^2/T$. 
The quadratic dependence  on $F_{\rm act}^2$  is similar to what was found for interacting active 
point-like particles~\cite{cugl-mossa1,cugl-mossa3} and interacting active polymers~\cite{cugl-mossa2,cugl-mossa3}
at low density, comparable to $\phi\simeq 0.1$ in our case where no aggregation 
effects exist. The square power-law dependence on $F_{\rm act}$  also applies to the dumbbell data for $\phi=0.1$.
Although in this double logarithmic scale the data seem to be very close to this form for all $\phi$, as we will see,
a closer look at them shows a non-trivial $\phi$ dependence. 

The results for $T_{\rm eff}$ for two working temperatures, $T=0.05$ and $T=0.1$,  are summarised in Table~\ref{tab:template}. 
At very small active force we find 
$T_{\rm eff} \simeq T$ at all densities, as expected as the system is near equilibrium (see the first rows in the two 
sets of data shown in Table~\ref{tab:template}). For intermediate
active forces, e.g. $F_{\rm act} =0.1$ (third rows in the tables), we see that $T_{\rm eff}$ is significantly larger than $T$ and 
that it weakly increases with increasing $\phi$. For still larger active forces the dependence on $\phi$ changes, as 
$T_{\rm eff}$ decreases with $\phi$ for  sufficiently large values of $\phi$ ($\phi \stackrel{>}{\sim} 0.1$ for $T=0.05$ and 
$\phi \stackrel{>}{\sim} 0.3$ for $T=0.1$, when $F_{\rm act} = 1$).
Consistently with what discussed in the previous paragraph, for fixed density, 
$T_{\rm eff}$ increases with $F_{\rm act}$ for the two working temperatures. Moreover, at fixed density and active force 
$F_{\rm act}>0.5$ we see that $T_{\rm eff}$ is lower at higher temperature.

\begin{table}[h]
\centering
\begin{tabular}{|l|c|c|c|c|}
\hline
 \multicolumn{5}{|c|}{$T=0.05$} \\
\hline  
 \  $F_{\rm act} \ $ & $\ \phi=0\ $ &	$\ \phi=0.1\ $   &	$\ \phi=0.3 \ $  &	$\ \phi=0.5 \ $ \\
 \hline
\ 0.001  & 0.0500	 & 0.0499       &	0.0488    &	0.0502 \\
\ 0.01  & 0.0509	&  0.0498       &	0.0502    & 	0.0522 \\
\ 0.1   & 0.142 	& 0.152	     &	0.179     &	0.223 \\
\ 0.5   & 2.35 & 2.51	     &	2.59	  &	2.44 \\
\ 1     & 9.27	& 9.68	     &	7.83       &	6.56 \\
\hline
\end{tabular}
\begin{tabular}{|l|c|c|c|c|}
\hline
 \multicolumn{5}{|c|}{$T=0.1$} \\
\hline  
 \  $F_{\rm act} \ $ & $\ \phi=0\ $ &	$\ \phi=0.1\ $   &	$\ \phi=0.3 \ $  &	$\ \phi=0.5 \ $ \\
 \hline
\ 0.001 & 0.100 & 0.102 & 0.100 & 0.097 \\
\ 0.01 & 0.100  & 0.100 & 0.100 & 0.101 \\
\ 0.1 & 0.146 & 0.158 & 0.167 & 0.201 \\
\ 0.5 & 1.25 & 1.34 & 1.56 & 1.77 \\
\ 1 & 4.71 & 5.03 & 5.10 & 4.92 \\
\hline
\end{tabular}
\caption{Values of $T_{\rm eff}$ from the analysis at different active forces and system's densities, at 
$T=0.05$ and $T=0.1$.
}
\label{tab:template}
\end{table}

\begin{figure}[h]
\begin{center}
  \begin{tabular}{cc}
%    \includegraphics[scale=0.9]{FIGS/figure_lavoro_Antonio/fig15/rapporto_chi_diff/values_of_T_eff_from_dumb.eps}
%    \\
    \includegraphics[scale=0.9]{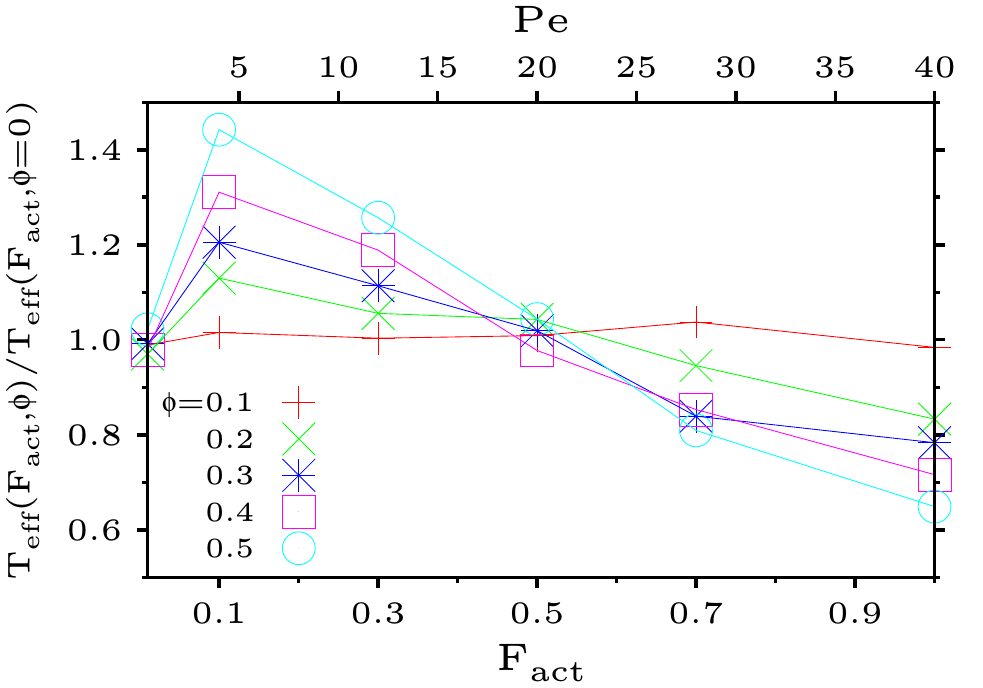}
  \end{tabular}
\caption{(Color online.) 
The ratio between the finite density effective temperature and the single molecule limit one as a function of the
strength of the active force,
$T_{\rm eff}(F_{\rm act}, \phi)/T_{\rm eff}(F_{\rm act}, 0)$,  where 
$T_{\rm eff}(F_{\rm act}, 0) = T \ [1+ \mbox{Pe}^2/8]$.   The temperature is $T=0.05$.
See the text for a discussion. 
}
    \label{T_eff_parametricplot2}
  \end{center}
\end{figure}

In Fig.~\ref{T_eff_parametricplot2} we display the ratio between the finite density effective temperature and the single molecule limit one as a function of the
strength of the active force, $T_{\rm eff}(F_{\rm act}, \phi)/T_{\rm eff}(F_{\rm act}, 0)$ {\it vs.} $F_{\rm act}$. 
If we could separate the diffusion coefficient of a single dumbbell from the $\phi$ dependence as 
proposed in Eq.~(\ref{diff_different_fi_approx}),  and if we had the same density dependence of the $\mu(F_{\rm act},\phi)$
as in the remaining factor in $D_A(F_{\rm act}, \phi)$, meaning $b(F_{act})=c(F_{act})$, then 
$T_{\rm eff}(F_{\rm act}, \phi)$ should be independent of $\phi$ and just equal to 
 $T_{\rm eff}(F_{\rm act}, 0)$. The data show that this holds 
for $\phi=0.1$ where the data points are (within numerical accuracy) constant but it does not for 
higher densities. For $\phi \ge 0.1$ one observes a non--monotonic dependence
of $T_{\rm eff}(F_{\rm act}, \phi)/T_{\rm eff}(F_{\rm act}, 0)$ on $F_{\rm act}$, reflecting in some way 
the non--monotonic behavior first shown for the diffusion and response ratios $D_A(F_{\rm act}, \phi)/D_A(F_{\rm act}, 0)$ and $\mu(F_{\rm act},\phi)/\mu(F_{\rm act}, 0)$. 
The finite density effective temperature is higher than the single molecule one for active forces 
in the interval $[0, 0.5]$ while it is lower than the single molecule one for active forces in the 
interval $[0.5, 1]$.  A maximum is reached at around $F_{\rm act} \simeq 0.1$. 
We ascribe the change in behaviour to the presence of finite size clusters in the 
sample for $F_{\rm act} \stackrel{>}{\sim} 0.5$ and sufficiently large density. These clusters 
(see the last panels in Fig.~\ref{fig:clusters})
would respond differently from the homogeneous bulk and their dynamics will be different.
 We also note that these curves seem to cross at the active force strength value $F_{\rm act}=0.5$.
A similar analysis of this ratio at a different temperature, $T=0.1$, shows that the crossing 
occurs at a different value of the active force strength,  $F_{\rm act} \simeq 1$,  that corresponds to 
the same P\'eclet number $\rm {Pe} = 20 $ (within numerical accuracy).

Finally, we note that $T_{\rm eff} >T_{\rm kin}$ as soon as the active force is intense enough to see a 
considerable deviation of both from the ambient temperature. Confront, for instance, the analytic values $T_{\rm eff} =0.142$
 and $T_{\rm kin}\simeq 0.050$ at $\phi=0$ and $F_{\rm act}=0.1$, and the numeric values $T_{\rm eff} \simeq 9.68$ and 
 $T_{\rm kin}\simeq 0.06$ at $\phi=0.1$ and $F_{\rm act}=1$.

\section{Conclusions}
\label{sec:conclusions}

In this paper we studied the dynamic properties of isolated and interacting, passive and active,  
dumbbell systems. 

The model studied here differs from other models of active 
matter previously studied in the literature with numerical techniques 
from a similar viewpoint. For instance,  we do not impose polar alignment mechanisms
as in~\cite{szabo,Henkes11}.
The active forces used here are different from the ones used in the 
analysis of active macromolecules presented in~\cite{cugl-mossa2,cugl-mossa3}:
while in the dumbbell system the active forces act along the main axis of the 
molecule, in the polymer system the active forces act only on the center monomer and 
in random directions. The Langevin  dynamics used in this paper 
is different from the Monte Carlo rule chosen in~\cite{Levis-Berthier} and 
from the run-and-tumble bacteria models in~\cite{Tailleur08,Tailleur09}.
Moreover, we did not distinguish translation and rotational noise, as done in~\cite{Fily12},
but we simply added independent Gaussian white noise (related to the friction dissipative term in the 
usual way) to the Langevin equations for the positions of the two atoms in the dumbbell.

For fixed temperature and active force strength, this very simple  model has a phase transition
between a homogeneous and a phase separated phase~\cite{Suma13,Suma14}. In this paper we
focused on the low density phase in which the system is homogeneous 
on average with, possibly, giant density fluctuations for sufficiently high density~\cite{Suma14}. 
We studied the dynamics for three different temperatures and 
a wide range of densities and active forces in this phase.

We first presented a detailed study of the mean-square displacement of the interacting active sample. 
We analysed the single passive and active dumbbell dynamics and we investigated how the 
finite density affects the various dynamic regimes and, in particular, the diffusive properties in the 
late time-delay limit that goes beyond the angular diffusion time of the single molecule. As one 
had expected, we found that the diffusion constant decreases with increasing density and increases
with increasing active force. The  ratio between the finite-density and the single particle 
diffusion constant exhibits an intriguing non-monotonic dependence on the active force.
We also analysed how the Tokuyama-Oppenheim density-dependence expression for 
this ratio~\cite{Tokuyama} is modified by self-propulsion, finding that a simpler exponential decay 
(with a non-monotonic active force dependent factor in the exponential) fits the data reasonably well.

We then studied the linear response function of the dumbbell displacement 
to infinitesimal perturbations that push them in random directions. As usual, to minimise the 
numerical error, we focused on the linear response integrated over  time
(instead of the instantaneous response that fluctuates much more). We found that, at fixed time delay, the 
integrated linear response decreases monotonically with increasing density. The dependence 
on the active force is again non-trivial, with a maximum response reached for active forces with 
strength in $0.1 \lesssim  F_{\rm act} \lesssim   0.3$ for densities in $0.1 \lesssim 
\phi \lesssim  0.3$.

The kinetic temperature is extracted from the asymptotic value of a one-time observable, the kinetic energy. As already discussed in detail 
in the context of glassy systems and granular matter~\cite{cugl-kur-pel,cugl:review}, although one proves that a system is not in equilibrium with  the thermal 
bath whenever $T_{\rm kin} \neq T$, the reverse is not true. In glassy systems in relaxation $T_{\rm kin}=T$ while these systems 
evolve out of equilibrium. The reason is that the kinetic temperature carries information about a fast observable, the kinetic energy, 
that can be able to quickly equilibrate with the environment, while other observables, being slower, can still be far from equilibrium.
Indeed, the kinetic temperature does not characterise  the large scale structural relaxation of glassy or driven systems.

The effective temperature notion, as obtained from the deviation from the equilibrium 
fluctuation-dissipation theorem linking spontaneous and induced fluctuations, has proven to be
very useful to understand the dynamics of slowly relaxing passive systems, such as 
glasses and gently sheared super-cooled liquids~\cite{cugl:review}. This concept has been explored in
some active systems as well as we explained in the introduction. In this work we studied 
the fluctuation-dissipation ratio for the low-density active dumbbell system and we characterised it
as a function of density and activity. The effective temperature is always higher than the ambient temperature, it increases with 
increasing activity and, for small active force it  monotonically increases with density while for sufficiently high activity 
it first increases to next decrease with the packing fraction. This effect should be due to the existence of finite-size 
clusters for sufficiently high activity and density at the fixed (low) temperatures at which we worked. The crossover occurs 
at lower activity or density the lower the external temperature. The finite density effective temperature is higher (lower) than
the single dumbbell one below (above) a cross-over value of the P\'eclet number. 

In the active dumbbell system we measured $T_{\rm kin}$ values that are very close to $T$ for 
$F_{\rm act}=0.01, \ 0.1, \ 1$.
The existence of the kinetic temperature characterising the short-time delay behaviour of 
fluctuations, and the effective temperature extracted from the fluctuation-dissipation relations at long time-delays, 
does not invalidate the possible thermodynamic interpretation of the latter. One simply has to focus on the 
dynamics of the systems in one or another dynamic regime~\cite{cugl-kur-pel,cugl:review} and test its thermodynamic properties within in it.

We want to stress once again that the effective temperature is not a parameter that characterises the 
statistical properties of the activity but an intensive parameter that tells us about the dynamic properties
of the interacting and active many-body system.

Fily and Marchetti~\cite{Fily12} argue that the effective tempeature notion cannot apply to active 
matter systems in their dense phase. They base their claim on the comparison of instantaneous 
snapshots of typical clustered configurations and equivalent thermal equilibrium ones
with similar overlap between particles at the same packing fraction. This argument cannot 
be used to refute the effective temperature ideas in this context (nor in 
the glassy context either) as its very definition is {\it dynamic} and correlation and linear response 
functions at {\it different} times need to be calculated to derive $T_{\rm eff}$. Having said this, 
the analysis of $T_{\rm eff}$ in a dense active phase has not been performed yet and we cannot 
claim that the concept will still be valid in this clustered phase.

The dumbbell system is particularly interesting as it allows one to study translational and, simultaneously, 
also rotational degrees of 
freedom. A recent analysis of the effective temperature ideas in a passive dumbbell system has shown that 
the relation between the two can depend non-trivially on the elongation of 
the molecule~\cite{Saverio14}. It would be very interesting to explore the effect of this 
parameter in the results that we showed and to investigate the fluctuation-dissipation relations 
of rotational degrees of freedom in the active dumbbell system as well.

The values of the effective temperatures found from the fluctuation-dissipation relation 
of the active system should be confirmed with alternative measurements to test their thermodynamic meaning.
For instance, one could use spherical passive particles as tracers and measure the 
effective temperature from their diffusion properties (as in~\cite{wu}) and kinetic energy fluctuations. 
We will report on the use of tracers in this context in a separate publication.

Finally, it would be interesting to analyse the effect that external potentials may have on the 
dynamics of this active dumbbell system. Tailleur and Cates derived a diffusive approximation 
for active run-and-tumble particles~\cite{Tailleur08,Tailleur09} and found that, for weak external potentials that modify only slightly the 
velocity of the particles, the system 
should follow equilibrium dynamics with the equilibrium bath temperature replaced by 
$T_{\rm eff}$. It will be interesting to check whether this result holds for the active dumbbell 
system under appropriate external forces and how it is modified by the finite density and
interactions within the sample, and strong external potential forces.

%\bibliographystyle{apsrev}
%\bibliography{dumbbells-biblio}

%\end{document}

\vspace{1cm}

\noindent Acknowledgments:
We thank  D. Loi and S. Mossa for very useful discussions. We warmly thank J. Tailleur for very useful comments 
on the manuscript. LFC is a member of Institut Universitaire de France. GG acknowledges the support of  
MIUR (project PRIN 2012NNRKAF).

\bibliographystyle{apsrev}
\bibliography{dumbbells}

\end{document}